\documentclass{article}

\usepackage{arxiv}
\usepackage[utf8]{inputenc} 
\usepackage[T1]{fontenc}    
\usepackage{hyperref}       
\usepackage{url}            
\usepackage{booktabs}       
\usepackage{amsfonts}       
\usepackage{nicefrac}       
\usepackage{microtype}      
\usepackage{lipsum}		
\usepackage{graphicx}
\usepackage{natbib}
\usepackage{doi}

\usepackage{amsmath}               
  \allowdisplaybreaks[1]           
\usepackage{amssymb}               
\usepackage{url}                   
\usepackage{rotating}              
\usepackage{multirow}              
\usepackage{lscape}                
\usepackage{tabularx}              
\usepackage{verbatim}              
\usepackage{footnote}              
\usepackage{float}                 
\usepackage{booktabs}              
\usepackage[base]{babel}           
\usepackage{subcaption}            
\usepackage{siunitx}               
\usepackage{indentfirst} 
\usepackage{ragged2e} 
\usepackage{caption} 
\usepackage{nameref}
\usepackage{xcolor}
\usepackage[normalem]{ulem} 

\title{Dose-Insensitive Defect Engineering, Carrier Kinetics, and Reproducible Chromaticity Tuning in Ion-Implanted InGaN/GaN Quantum Wells}

\date{} 					

\author{Quan-Shan Liu\\
	Department of Electrical and Electronic Engineering\\
	Photon Science Institute, University of Manchester\\
	Oxford Road, Manchester, M13 9PL, UK \\
	\texttt{quan-shan.liu@manchester.ac.uk} \\
    \And
	Mason Adshead\\
	Department of Electrical and Electronic Engineering \\
    Photon Science Institute, University of Manchester \\
	Oxford Road, Manchester, M13 9PL, UK \\
	\texttt{mason.adshead@manchester.ac.uk} \\
	\And
	Sadia Sheraz\\
	Department of Chemistry\\
	Photon Science Institute, University of Manchester\\
	Oxford Road, Manchester, M13 9PL, UK \\
	\texttt{sadia.sheraz@manchester.ac.uk} \\
	\And
	Maddison Coke\\
	Department of Electrical and Electronic Engineering \\
    Photon Science Institute, University of Manchester \\
	Oxford Road, Manchester, M13 9PL, UK \\
	\texttt{maddison.coke@manchester.ac.uk} \\
	\And
	Nicholas Lockyer\\
	Department of Chemistry \\
    Photon Science Institute, University of Manchester \\
	Oxford Road, Manchester, M13 9PL, UK \\
	\texttt{nick.lockyer@manchester.ac.uk} \\
	\And
	Richard J. Curry \thanks{Author to whom any correspondence should be addressed.} \\
	Department of Electrical and Electronic Engineering \\
    Photon Science Institute, University of Manchester\\
	Oxford Road, Manchester, M13 9PL, UK \\
	\texttt{richard.curry@manchester.ac.uk} \\
}

\renewcommand{\shorttitle}{Dose-Insensitive Defect Engineering, Carrier Kinetics, and Reproducible Chromaticity Tuning in Ion-Implanted InGaN/GaN Quantum Wells}

\hypersetup{
pdftitle={A template for the arxiv style},
pdfsubject={q-bio.NC, q-bio.QM},
pdfauthor={David S.~Hippocampus, Elias D.~Striatum},
pdfkeywords={First keyword, Second keyword, More},
}

\begin{document}
\maketitle

\begin{abstract}
Post-growth defect engineering via ion implantation provides a powerful pathway for spatial optical profiling and colour patterning in III-V alloying photonic integration. However, the comprehensive recombination kinetics governing deep-level defect saturation and excitonic recovery under high-temperature annealing remain insufficiently understood. Here, we present a systematic study on the optical dynamics, rate-equation kinetics, and chromaticity evolution of indium-implanted InGaN/GaN multiple quantum wells (MQWs) across implantation doses (5E14-5E16 ions cm$^{-2}$) and subsequent thermal annealing stages (500-1100 °C). Photoluminescence (PL) spectrum analysis reveals that a dose of $\le$5E14 ions cm$^{-2}$ induces a modification of the optical response that does not further change upon high-dose implantation. A two-channel coupled rate-equation model is fitted to the data, optimised via differential evolution, allowing the extraction of transition rate constants. This demonstrates that thermal processing at 1000 °C suppresses the carrier capture rate into deep-level states and also reduces its radiative recombination rate. This kinetic bottleneck drives an order-of-magnitude extension in the channel-specific radiative lifetime. Leveraging the excitation power density dependence of the differential recombination kinetics, where deep defects saturate whilst MQW emission scales near-linearly, we achieve a universal and power density-tunable chromaticity trajectory from warm yellow to cool white-blue emission. These insights enable microscopic defect physics to be linked to the macro-scale colour tailoring observed.
\end{abstract}


\section{Introduction}
\label{Section1}

Monolithic optoelectronic integration based on III-V compound semiconductors has emerged as a cornerstone technology for next-generation light-emitting diodes and laser diodes\citep{nakamura1994candela,nakamura1998roles,wang2017wavelength,fan2023monolithically}. Crucial to these emitting devices, InGaN/GaN MQWs offer highly tunable direct bandgaps spanning the entire visible spectrum\citep{vurgaftman2001band,vurgaftman2003band}. However, though traditional methods like metal-organic chemical vapour deposition and molecular beam epitaxy exhibit superior advantages in growing high-quality epilayers\citep{capper2017epitaxial,wang2004molecular}, achieving spatial multi-colour pixelation or localised optical phase/intensity profiling on a single substrate remains technologically challenging. Consequently, post-growth defect engineering techniques capable of modifying optical properties with sub-micron spatial resolution are highly sought after.

Ion implantation represents a versatile, highly reproducible, and spatially precise strategy for local material modification\citep{jacobson1995high}. By delivering energetic ions into the crystal lattice, atomic displacements generate controlled point defect networks that modify local non-radiative and radiative carrier recombination paths. Notably, the choice of implanted species is paramount. Whilst noble gases or transition metals often introduce foreign chemical impurities\citep{langer2013nonradiative,wang2020room}, and elements like magnesium or silicon induce unwanted electrical doping\citep{kusakabe2005impurity}, indium (In) ions serve as a strictly isoelectronic, constituent species for InGaN/GaN heterostructures. The use of In ions preserves the intrinsic chemical stoichiometry while leveraging their heavy atomic mass to efficiently generate localised displacement cascades without introducing extraneous chemical contaminants or shallow donor/acceptor states.

Despite extensive studies on ion-implantation-induced damage in bulk GaN\citep{ronning2001ion,kucheyev2001ion}, a comprehensive understanding of carrier recombination kinetics within thin InGaN/GaN MQWs under extreme post-implantation thermal budgets remains incomplete. In this work, we present a systematic study on the optical properties, including recombination dynamics, and chromaticity evolution, of indium-implanted InGaN/GaN MQWs subjected to rapid thermal annealing (RTA) from 500 °C to 1100 °C. Through excitation-dependent photoluminescence (PL) characterisation coupled with a two-channel rate-equation model, we track the competition between emission bands across implanted In doses from 5E14 to 5E16 ions cm$^{-2}$ upon 500 °C and 1000 °C annealing. Furthermore, we map the microscopic defect dynamics directly onto macroscopic chromaticity space. It is demonstrated that while high-temperature thermal processing at 1000 °C does not fully recover the implantation-induced damage, a universal and power density-tunable chromaticity trajectory from warm yellow to cool white-blue emission is possible. These findings provide critical kinetic insights into post-growth chromaticity tailoring in monolithic III-V compound photonic devices.

\section{Characterisation methods}
\label{Section2}
The sample used in this study is an as-received specimen described in our previous work\citep{liu2026thermal}. Briefly, the sample contains a wurtzite GaN template grown on a sapphire substrate, with a buried five-layer InGaN/GaN MQW architecture. The MQW structure was estimated to have an average period of 11.68 ±0.54 nm and a nominal indium molar fraction of approximately 15\%–18\%. To investigate the effect of ion implantation, selected areas of the sample were subjected to 25 keV indium ion implantation, while leaving other regions unimplanted for control experiments. There are 15 50$\times$50 \textmu m$^{2}$ implanted squares in total, distributed in a 3 (row) by 5 (column) array. Between adjacent implanted square regions, the horizontal and vertical separations are 100 and 200 \textmu m, respectively. The five implanted regions in each row represent an increasing dose of 5E14, 1E15, 5E15, 1E16, and 5E16 ions cm$^{-2}$ from left to right. Each of the three rows is the same so that each column provides three regions with identical dose for study.

The ion implantation was performed using the Platform for Nanoscale Advanced Materials Engineering (P-NAME) Facility (Ionoptika, Q-One) at the University of Manchester\citep{adshead2023high}. Post-implantation thermal annealing is typically required for both repairing ion damage and activation of implanted dopants\citep{mayer1970eriksson}. Prior to annealing an AlN capping layer was deposited to prevent the GaN from potential decomposition and implanted indium from escaping\citep{zolper1996sputtered,cao1998ultrahigh}. The capping layer thickness was estimated to be $\sim$ 55.67 ±1.67 nm via atomic force microscopy (AFM)\citep{liu2026thermal}. Subsequently, annealing was carried out under a nitrogen atmosphere using an AnnealSys As-One 100 rapid thermal processor, with a controlled ramp rate of 1 °C/s. Sample cooling was aided by a circulating water-cooling system. The sample annealing temperature increased in a stepwise manner from 500 °C to 1100 °C in 100 °C increments, each held at the maximum temperature for 60 seconds. Following each anneal the sample's optical properties were characterised before proceeding with subsequent annealing, summarised in Table \ref{table:stage definition} for clarification. The same sample underwent each successive annealing stage, so the total thermal budget delivered over the course of the study is cumulative.

    \begin{table}[!ht]
        \centering
        \caption{Summary of annealing cycles undertaken.}
        \begin{tabular}{ccccc}
        \hline
            Stage & Definition & Comments  \\ \hline
            1 & Uncapped & selected areas subject to In implantation  \\ 
            2 & Capped & deposited with $\sim$55.67 nm AlN \\
            3 & Post-annealed at 500 °C & N$_2$ atmosphere for 60 seconds \\
            4 & Post-annealed at 600 °C & N$_2$ atmosphere for 60 seconds \\
            5 & Post-annealed at 700 °C & N$_2$ atmosphere for 60 seconds \\
            6 & Post-annealed at 800 °C & N$_2$ atmosphere for 60 seconds \\
            7 & Post-annealed at 900 °C & N$_2$ atmosphere for 60 seconds \\
            8 & Post-annealed at 1000 °C & N$_2$ atmosphere for 60 seconds \\
            9 & Post-annealed at 1100 °C & N$_2$ atmosphere for 60 seconds \\
            \hline
        \end{tabular}
        \label{table:stage definition}
    \end{table}

PL intensity mapping was performed for stages 1 to 3 using a Raman spectroscopy microscope (incorporating a Kimmon IK series He-Cd 325 nm laser and a Horiba LabRAM HR Evolution spectrometer). The laser beam was focused onto a $\sim$20 \textmu m spot with the use of a 15X near-ultraviolet objective lens. PL mode was activated by switching the built-in measurement unit from cm$^{-1}$ to nm. Spatial PL maps were obtained by scanning using the following parameters: step resolution 25 \textmu m, acquisition time 0.01 s, single scan accumulation, wavelength range 350-600 nm, and grating groove density 1800 gr/mm. The intensity of each site mapped was obtained by calculating the mean intensity value across the overall specified wavelength range.

Steady-state PL spectra were obtained for all stages summarised in Table \ref{table:stage definition}. The sample was excited by the same 325 nm laser, and the resulting emission was collected by a Thorlabs fibre-coupled spectrometer (CCS200). The collected spectra were corrected for the wavelength-dependent responsivity of the system and the AlN capping layer transmittance (as described in \citep{liu2026thermal}). Integrated intensities of main emission bands were extracted for data analysis. Acquisition of the integrated intensity progressed by obtaining the product of corrected PL intensity $I^{PL}(\lambda)$ and $\lambda^2$ to plot the spectra as a function of photon energy\citep{reshchikov2018two,pelant2012luminescence}. Each PL peak observed was separated for Gaussian fitting, with intensity integration of each fitted peak then performed.

Additional excitation power density-dependent PL spectra were recorded, using neutral-density (ND) filters to attenuate the excitation power, for the sample post-annealing at 500 °C and 1000 °C. The laser power density was varied from ~2021 $\text{W/cm}^2$ (no attenuation) to 18.30 $\text{W/cm}^2$. The corrected intensity $I^{PL}(\lambda)$ was again converted by multiplying by $\lambda^2$ before Gaussian fitting and intensity integration. All PL-related characterisation was completed at room temperature.

Rate equation modelling was subsequently executed to fit the excitation power density-dependent PL behaviour. Details of the model and the calculation principles are reported elsewhere\citep{liu2026thermal}. The excitation power density-dependent PL intensity $I^{PL}(\lambda)$ underwent similar data processing, including unit conversion, Gaussian fitting and intensity integration. The only difference is that $\lambda^{3}$ instead of $\lambda^{2}$ was multiplied to give y values proportional to photon numbers as a function of photon energy\citep{reshchikov2018two,pelant2012luminescence}. 

The Commission Internationale de l’Eclairage (CIE) 1931 chromaticity coordinates were obtained to reveal the colour shift of the sample as annealing proceeded and as a function of excitation power density. By numerically integrating the corrected PL spectrum $I^{PL}(\lambda)$ with the CIE 1931 2° colour matching functions, the tristimulus values $(X, Y, Z)$ were derived. Next, the chromaticity coordinates $(x, y)$ were determined using standard normalisation formulas: $x = X/(X+Y+Z)$ and $y = Y/(X+Y+Z)$.

Finally, to investigate the sample's spatial composition post-annealing, we performed ToF-SIMS using a J105 SIMS instrument (Ionoptika Ltd, Chandlers Ford, U.K.). In positive ion mode, a focused C$^{+}_{60}$ cluster primary ion source was rastered over a 100 $\times$ 100 \textmu m$^2$ area with a 2 \textmu m step size. We maintained the ion dose per layer at $\sim$1.13E13 ions/cm$^2$ to optimise both the sputtering rate and depth resolution. To ensure a flat crater bottom and eliminate edge artefacts, we integrated the analytical data only from a central region of 25$\times$25 pixels obtained from the centre of the 50 $\times$ 50 \textmu m$^2$ sputtered zone covering the implantation site.

In the relevant figures, unless specified otherwise, the vertical error bars show the uncertainty propagated from the fitting covariance matrix. The horizontal error bars reflect conservative estimates based on the digital resolution of the recorded laser power data. Results of the unimplanted control samples are adapted from previously reported work\citep{liu2026thermal}.

\section{Results and discussion}
\label{Section3}

\subsection{Steady-state photoluminescence spectrum analysis}
\begin{figure}[htbp]
      \centering
      \includegraphics[width=1\textwidth]{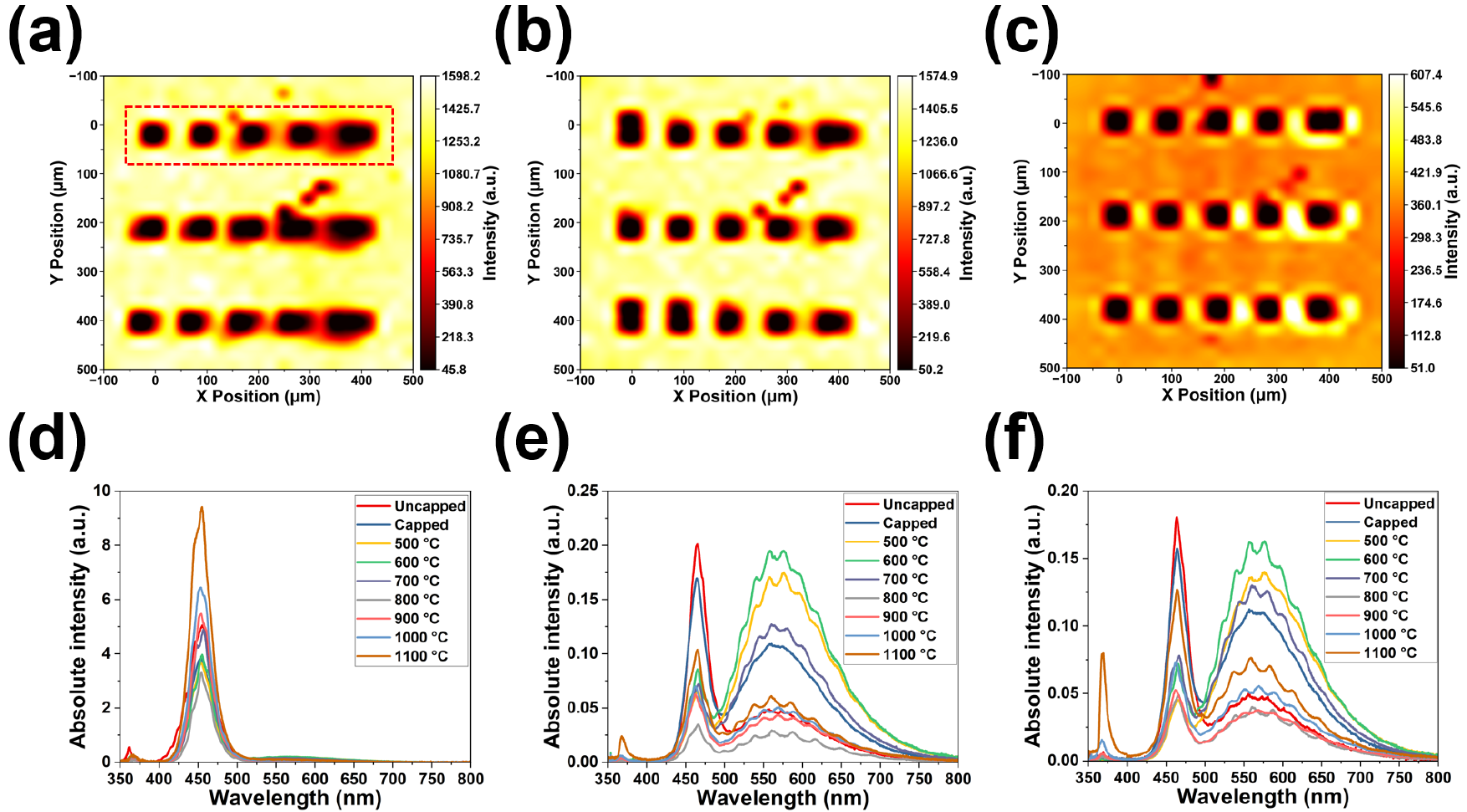}
      \caption{Room temperature PL mapping of the implanted areas at (a) post In implantation (stage 1), (b) post AlN capping (stage 2), and (c) post 500 °C thermal annealing (stage 3). The red dashed box indicates the specific implanted regions utilised for the further characterisation reported within. Room temperature PL spectra obtained following each stage of processing from (d) an unimplanted region, (e) the 5E14 ions cm$^{-2}$ implanted region, and (f) the 5E16 ions cm$^{-2}$ implanted region.}
      \label{figure_mapping}
    \end{figure}

Figures \ref{figure_mapping}(a) to \ref{figure_mapping}(c) display the PL intensity maps of the 3 by 5 implanted micro-array at experimental stages 1 to 3 (Table \ref{table:stage definition}). The spatial mapping clearly demonstrates high-contrast optical patterning defined by the implanted square micro-arrays, presenting horizontally increasing doses from left to right (5E14, 1E15, 5E15, 1E16, and 5E16 ions cm$^{-2}$ respectively) and vertical repetition. Upon 25 keV indium ion implantation severe PL quenching is observed in the implanted regions owing to the introduction of ion-induced damage and non-radiative recombination centres. The impact of ion implantation increases with In dose across each row, suggesting accumulation of damage. The deposition of the thin AlN capping layer (stage 2) does not have any noticeable effect on the PL intensity. However, there is an overall significant reduction in PL intensity across the entire sample after 500 °C thermal annealing. This is mainly attributed to the quenching of MQW emission (vide infra). Each of the three repeated rows displays stable spatial uniformity and intensity consistency. This robust inter-site reproducibility justifies isolating the first row of implanted regions (highlighted by the red dashed box in Figure \ref{figure_mapping}(a)) for subsequent optical characterisation.

The spectral evolution across progressive thermal annealing stages is detailed in Figures \ref{figure_mapping}(d) to \ref{figure_mapping}(f) for an unimplanted control region, the 5E14 ions cm$^{-2}$, and the 5E16 ions cm$^{-2}$ implanted region respectively. The PL spectra obtained from regions with other implanted doses can be found in Supplementary Information, section 1. The overall luminescence is composed of three primary emission bands: a near-band-edge ultraviolet band (UVB, $\sim$363 nm) originating from the wurtzite GaN template\citep{kim1997multiphoton}, a prominent blue band (BB, $\sim$455 nm) associated with radiative exciton recombination inside the InGaN/GaN MQWs, and a broad yellow band (YB, $\sim$565 nm) on the low-energy side of the MQW emission, attributed to deep-level defects including C$_{N}$\citep{reshchikov2014carbon}, C$_{N}$O$_{N}$\citep{demchenko2013yellow}, and V$_{Ga}$ complexes\citep{reshchikov2005luminescence,ogino1980mechanism,neugebauer1996gallium}. For the unimplanted control region, while annealing causes intensity variations of all PL bands, the dominance of BB emission has not been affected. Conversely, upon ion implantation, the relative peak intensity between BB and YB changes as each of the experimental stages proceeds. We note that in most cases the UVB is below the detection limit. For this reason, the UVB peak was excluded from the Gaussian peak fitting in the analysis within.

Inspection of the BB peak position indicates that a slight red-shift (from $\sim$455 nm to $\sim$462 nm) occurs after implantation. At implanted sites, a significant portion of photo-generated carriers are captured by implant-induced defects, leading to dominant non-radiative recombination and heat dissipation. Consequently, the steady-state carrier density within the InGaN/GaN MQWs is severely reduced. This depletion of carriers weakens the screening effect against the internal piezoelectric field\citep{takeuchi1997quantum,chichibu2002localized}, leaving the energy bands highly tilted. The result of this is the observed red shift of the BB emission peak due to the enhanced quantum-confined Stark effect (QCSE)\citep{bernardini1997spontaneous}. This mechanism is also supported by PL excitation power density behaviour, where decreasing the excitation power density yields a similar red-shift trend\citep{peng1999piezoelectric}. This indicates that the implanted sites behave similarly to the pristine sample under low-injection conditions where the polarisation field remains unshielded. Radiative transitions thus preferentially occupy lower energy states, including the YB and the long-wavelength tail of the BB.

    \begin{figure}[htbp]
      \centering
      \includegraphics[width=1\textwidth]{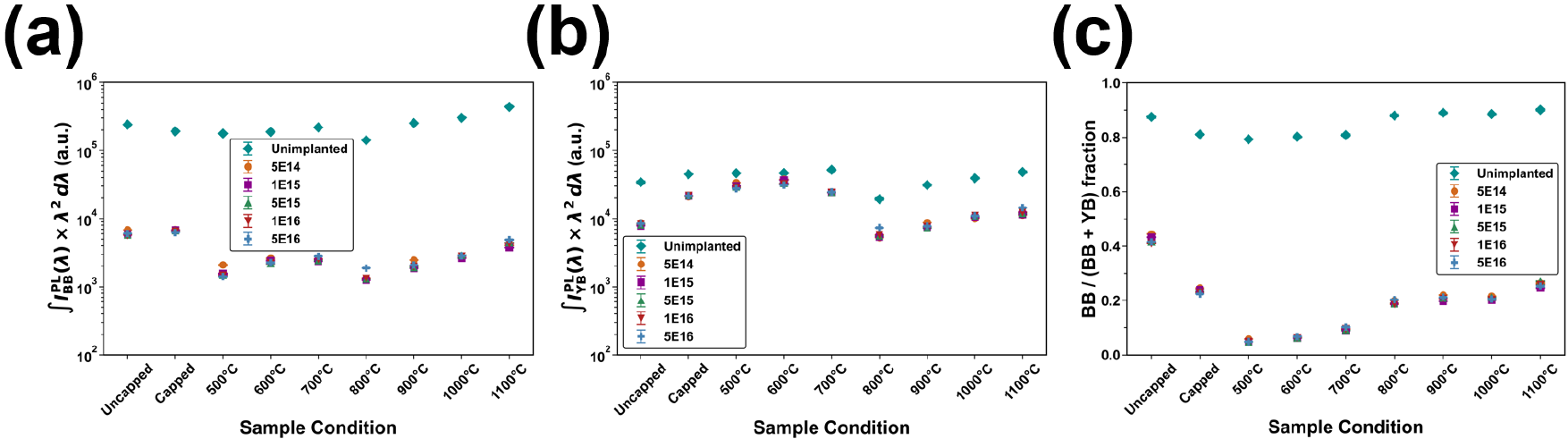}
      \caption{Integrated intensity of the BB PL (a) and YB PL (b) following each experimental stage for each implantation dose. (c) The changing fraction of BB PL obtained using the data in (a) and (b).}
      \label{Figure_intensity}
    \end{figure} 

Figures \ref{Figure_intensity}(a) and \ref{Figure_intensity}(b) present the integrated intensity of the BB and YB PL obtained after each experimental stage for each In-implanted dose. Likewise, Figure \ref{Figure_intensity}(c) displays how the relative strength of the BB PL changes with respect to the total (BB + YB) PL emission. For the unimplanted area, the BB dominates across all process stages, contributing a high fraction (80\%-90\%) of the total PL. In contrast, ion implantation dramatically quenches the BB intensity by over an order of magnitude and makes the YB PL prominent.

Upon annealing at temperatures increasing from 500 °C to 700 °C the YB PL intensity is seen to increase. This is accompanied by a more significant reduction in the BB PL intensity, leading to a minimum BB PL contribution to the total occurring following annealing at 500 °C. This behaviour may be attributed to the thermal activation and reconfiguration of mobile vacancies into stable deep-level complexes, which contribute to the YB emission while suppressing the BB emission. Upon further annealing at 800 °C both the BB and YB PL emission intensities are reduced, suggesting the transient formation of competitive non-radiative recombination centres which have previously been attributed to GaN surface decomposition\citep{king1998cleaning}. However, we observe that upon further annealing from 900 °C to 1100 °C partial recovery of both the BB and YB emission occurs. This implies that the AlN capping layer is providing the intended protection against decomposition.

A consistent observation shown by Figure \ref{Figure_intensity} is that the integrated intensity evolution of BB and YB PL is highly reproducible, independent of the In implantation dose. This strongly suggests that the optical damage saturates at or below a dose of 5E14 ions cm$^{-2}$, and that the post-implantation ratio of BB to YB PL is primarily governed by temperature-driven defect kinetics rather than the initial ion dose.

\subsection{Excitation-dependent photoluminescence spectrum analysis}
\begin{figure}[htbp]
      \centering
      \includegraphics[width=1\textwidth]{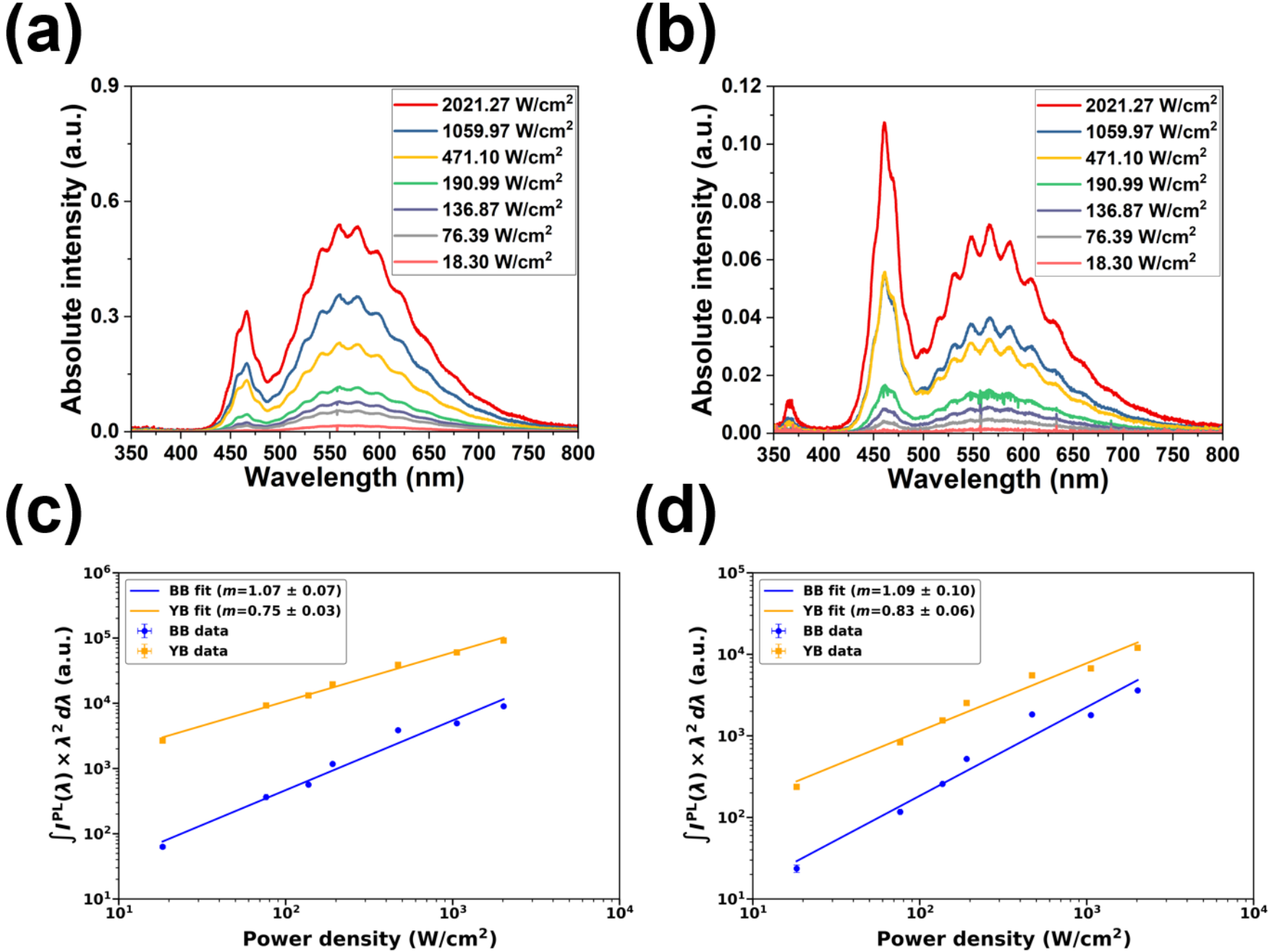}
      \caption{Excitation power density dependence of 325 nm excited PL obtained from a region implanted with an In dose of 5E15 ions cm$^{-2}$ obtained after (a) 500 °C and (b) 1000 °C annealing. A Log-Log plot of the integrated PL intensity as a function of excitation power density for the main emission peaks obtained after (c) 500 °C and (d) 1000 °C annealing.}
      \label{figure_PD_PL_5E15}
    \end{figure}

To further understand the recombination dynamics and defect saturation behaviour, excitation power density-dependent PL measurements were conducted. Figures \ref{figure_PD_PL_5E15}(a) and \ref{figure_PD_PL_5E15}(b) depict the excitation power density-dependent PL spectra of the 5E15 ions cm$^{-2}$ implanted region following 500 °C and 1000 °C annealing respectively. The results for other dose-implanted areas are provided in Supplementary Information, section 2. 

Following annealing at 500 °C, the broad YB dominates the PL spectra over the entire excitation power density range, indicating that deep-level defect complexes act as the primary radiative channels. After subsequent annealing at 1000 °C the overall absolute PL intensity is reduced, likely due to the activation of competitive non-radiative recombination centres and MQW interdiffusion. However, with increasing excitation power density the BB PL peak intensity increases at a faster rate than the YB PL counterpart and at the highest power becomes dominant. This power density-dependent spectral maximum inversion demonstrates that the higher-temperature thermal annealing has successfully suppressed deep-level defect recombination, allowing the intrinsic MQW emission to re-emerge as the competitive radiative recombination pathway under high injection levels. Furthermore, the UVB PL peak re-emerges, signifying limited structural restoration of the GaN matrix.

The integrated PL intensity ($\int I^{PL}(\lambda)\times \lambda^2 \, d\lambda$) as a function of excitation power density ($P$) is fitted using the power-law relationship of $\int I^{PL}(\lambda)\times \lambda^2 \, d\lambda \propto P^m$, where the exponent $m$ reflects the dominant carrier recombination pathway (Figures \ref{figure_PD_PL_5E15}(c) and \ref{figure_PD_PL_5E15}(d). For the BB PL, the fitted power-law exponent yields $m$ = 1.07 ±0.07 at 500 °C and $m$ = 1.09 ±0.10 at 1000 °C. A power exponent moderately exceeding unity ($m > 1$) signifies a dominant excitonic emission alongside the prominent role of Shockley-Read-Hall (SRH) non-radiative recombination centres introduced by ion bombardment\citep{schmidt1992excitation}. Under low photo-excitation densities, photogenerated carriers are preferentially captured by these unannealed SRH defect states. As the excitation density increases, the progressive saturation of these non-radiative channels leads to a super-linear rise in BB PL.

In sharp contrast, the YB exhibits pronounced sub-linear scaling with $m$ = 0.75 ±0.03 at 500 °C and $m$ = 0.83 ±0.06 at 1000 °C. The sub-linear behaviour ($m < 1$) provides evidence of a finite population of defect-mediated recombination \citep{schmidt1992excitation,reshchikov2005luminescence}. After annealing at 1000 °C the integrated intensity of the YB PL is reduced, signifying a reduced radiative recombination probability. The slight increase in $m$ observed following annealing indicates that the onset of saturation for the YB PL occurs at a higher excitation power density. The decreasing trend of the YB PL intensity is possibly driven by either the reduction of active deep-level defects following annealing or a proliferation of thermally propagated non-radiative defect networks. Concurrently, the comparatively robust BB recombination channel keeps draining photogenerated carriers and suppresses carrier accumulation at the remaining YB defect states, thereby elevating the optical power density threshold required to fill these trap states.

The power-law exponents ($m$) for the BB and YB PL excitation power density dependence are provided in Supplementary Information, section 2. Within the fitting uncertainty, there is no change in the values of $m$ with implanted dose following annealing at 500 °C, again suggesting that any ion implantation-induced damage saturates at or below a dose of 5E14 ions cm$^{-2}$.

After annealing at 1000 °C though some variation in $m$ is observed it is within the fitting uncertainty. The most significant observation is that this annealing seems to lead to similar behaviour for both the BB and YB PL excitation power density dependence in the unimplanted and implanted regions of the sample. This suggests that 1000 °C annealing provides sufficient thermal budget to heal point-like SRH centres in low-dose samples, eliminating the super-linear SRH competition and restoring intrinsic MQW recombination dynamics. For the unimplanted area, a power exponent remains near unity ($m \approx$ 1) which is consistent with standard excitonic recombination regardless of annealing temperature.

\subsection{Rate equation modelling analysis}

To quantitatively deconvolve the competing carrier recombination pathways across different implantation doses and thermal processing stages, we employ a two-channel steady-state rate equation framework previously established for the unimplanted region in \citep{liu2026thermal}. Briefly, given that the UVB and BB PL both exhibit a power-scaling exponent of $m \gtrsim$ 1 at the unimplanted area and that the UVB PL is orders of magnitude weaker than the other PL emission, the excitonic radiative channels are consolidated into a unified upper state (the combined UVB + BB reservoir). Starting from this upper state, the localised carriers follow three distinct pathways: direct radiative recombination producing UVB and BB, non-radiative SRH recombination, or irreversible trapping by a finite density of deep-level defect states which subsequently drive the saturable YB radiative transitions.

    \begin{equation}
    \begin{cases}
        \begin{aligned}
        dN_1/dt = G - \left( k_{r1} + k_{nr1} + k_{capture} \left( N_{defect} - N_2 \right) \right) N_1 = 0 \\ 
        dN_2/dt = k_{capture} N_1 \left( N_{defect} - N_2 \right) - \left( k_{r2} + k_{nr2} \right) N_2 = 0
        \end{aligned}
        \label{equ:rate equation}
    \end{cases}
    \end{equation}
    
As shown in Equation \ref{equ:rate equation}, two coupled rate equations under steady-state conditions have been constructed to parameterise the kinetics. Parameters and variables include: the excited carrier population at the UVB plus BB (YB) channel ($N_{1(2)}$) over time $t$, the photo-generation rate $G = \beta P$ proportional to power density $P$ via an equivalent scaling constant $\beta$, the radiative/non-radiative recombination rates of the UVB plus BB channel ($k_{r1/nr1}$), the radiative/non-radiative recombination rates of the YB channel ($k_{r2/nr2}$), the carrier capture rate from the UVB plus BB channel to the YB channel ($k_{capture}$), and the maximum permitted population of excited carrier at defect states ($N_{defect}$). 

Channel-specific metrics such as lifetime and radiative efficiency can be defined in Equations \ref{equ:emission lifetime} and \ref{equ:emission efficiency}. It should be noted that these parameters denote intrinsic values confined strictly to a designated recombination channel, distinct from the overall effective lifetime and quantum efficiency of the entire sample.

    \begin{align}
    \tau_1 &= \frac{1}{k_{r1}+k_{nr1}}, \qquad
    \tau_2 = \frac{1}{k_{r2}+k_{nr2}} \label{equ:emission lifetime}
    \end{align}

    \begin{align}
    \eta_{UVB+BB} &= \frac{k_{r1}}{k_{r1}+k_{nr1}}, \qquad
    \eta_{YB} = \frac{k_{r2}}{k_{r2}+k_{nr2}} \label{equ:emission efficiency}
    \end{align}

To prevent unphysical parameter fitting and to account for experimentally determined lifetimes ($\sim\text{ns}$ and $\sim\mu\text{s}$ for the BB and YB PL, respectively), the parameter space was global-optimised using a differential evolution scheme by fitting the modelled intensities ($\int I\mathrm{_{UVB+BB}^{Model}} \times \lambda^3 \, d\lambda = k_{r1}N_1$, $\int I\mathrm{_{YB}^{Model}}\times \lambda^3 \, d\lambda = k_{r2}N_2$) to the power density-dependent experimental datasets $\int I\mathrm{_{UVB+BB}^{Exp}}\times \lambda^3 \, d\lambda$ and $\int I\mathrm{_{YB}^{Exp}}\times \lambda^3 \, d\lambda$, regularised by a soft/hard window penalty function, as detailed in \citep{liu2026thermal}. The global optimisation was repeated five times with randomised initial conditions to derive the mean and standard deviation of the extracted parameters. The parameter set achieving the lowest total loss function across these runs was selected as the best-fit solution, and its $N_{1/2}$ values were used to finalise $\int I\mathrm{_{(UVB+BB)/YB}^{Model}}\times \lambda^3 \, d\lambda$.

    \begin{figure}[htbp]
      \centering
      \includegraphics[width=1\textwidth]{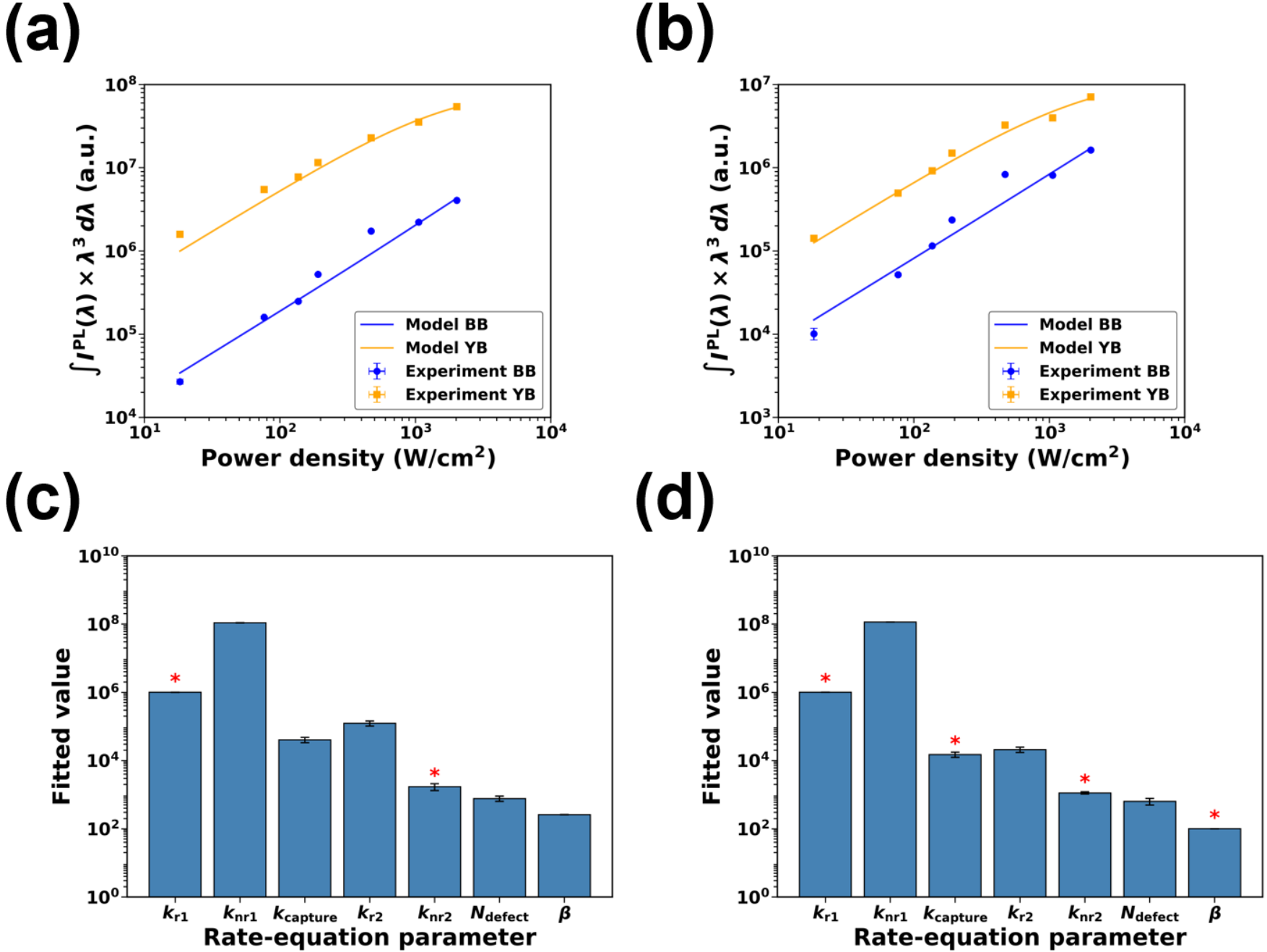}
      \caption{For the 5E15 ions cm$^{-2}$ implanted area. Comparison between the emission channels' experimental PL integrated intensities (data points) and the best fit obtained using the rate-equation model (lines) as a function of excitation power density after annealing at (a) 500 °C and (b) 1000 °C. Mean fitting parameter values and standard deviation obtained from repeated differential evolution runs after annealing at (c) 500 °C and (d) 1000 °C. An `*' identifies a parameter whose best-fit value lies within 5\% of a search-space boundary.}
      \label{figure_REM_5E15}
    \end{figure} 

Here, we extend this benchmarked model previously applied to unimplanted GaN/InGaN to evaluate the carrier dynamics of the implanted areas. Due to the post-implantation quenching of UVB, this peak was excluded from the Gaussian fitting during the subsequent rate equation modelling. Figure \ref{figure_REM_5E15} presents the best fit obtained using the model of the 5E15 ions cm$^{-2}$ implanted area following 500 °C and 1000 °C annealing. The results of fitting for other implanted doses can be found in Supplementary Information, section 3. As shown in Figures \ref{figure_REM_5E15}(a) and \ref{figure_REM_5E15}(b), the numerical solutions derived from the rate-equation model result in good fitting of the experimentally obtained PL integrated intensities across two orders of magnitude in power density, with the coefficient of determination ($R^2$) for both fitting curves greater than 0.90 (see Supplementary Information, section 3). The model successfully reproduces both the near-linear trend of the $\int I\mathrm{_{BB}^{Exp}}\times \lambda^3 \, d\lambda$ and the sub-linear saturation behaviour of the $\int I\mathrm{_{YB}^{Exp}}\times \lambda^3 \, d\lambda$ at 500 °C and 1000 °C. This strong agreement confirms that the underlying physical picture, comprising direct excitonic recombination competing with carrier capture into saturable deep defect states, accurately reflects the dynamic steady-state conditions in the implanted areas.

The carrier dynamic evolution, the fitted rate-equation parameters and the optimisation stability metrics are summarised in Figures \ref{figure_REM_5E15}(c) and \ref{figure_REM_5E15}(d). In alignment with our previous optimisation methodology\citep{liu2026thermal}, the error bars reflect the reproducibility of the differential evolution across independent runs rather than strict parameter confidence intervals. Furthermore, parameters flagged with asterisks denote values lying within $5\%$ of a search-space boundary (e.g., $k_{r1}$, $k_{nr2}$, $k_{capture}$ and $\beta$), indicating boundary-truncated solutions imposed by model constraints. Whilst the absolute magnitudes of individual parameters should not be over-interpreted as uniquely determined physical constants due to parameter degeneracies, their relative trends across sample conditions remain of use.

Comparison of the fitted parameter sets between 500 °C and 1000 °C is provided in Figure \ref{figure_REM_5E15}(c) and Figure \ref{figure_REM_5E15}(d). The carrier capture rate ($k_{capture}$) is seen to decrease from $\sim10^{4.6}\text{ s}^{-1}$ (500 °C) to $\sim10^{4.0}\text{ s}^{-1}$ (1000 °C), confirming a significantly lowered probability for photo-excited carriers to be captured into the deep-level trap network. Concurrently, the radiative recombination rate of the YB emission ($k_{r2}$) reduces by nearly an order of magnitude from $\sim10^{5.0}\text{ s}^{-1}$ (500 °C) to $\sim10^{4.1}\text{ s}^{-1}$ (1000 °C). As a result, the combined drop in both $k_{capture}$ and $k_{r2}$ limits the YB PL intensity, resulting in the BB emission dominating the PL spectra (Figure \ref{figure_PD_PL_5E15}(b)). Meanwhile, the equivalent scaling constant $\beta$ undergoes a reduction from $\sim10^{2.4}\text{ s}^{-1}$ (500 °C) to $\sim10^{2.0}\text{ s}^{-1}$ (1000 °C), indicating that a population reduction of photo-excited carriers may also be a reason for the overall intensity quenching observed at higher annealing temperatures.

    \begin{figure}[htbp]
      \centering
      \includegraphics[width=1\textwidth]{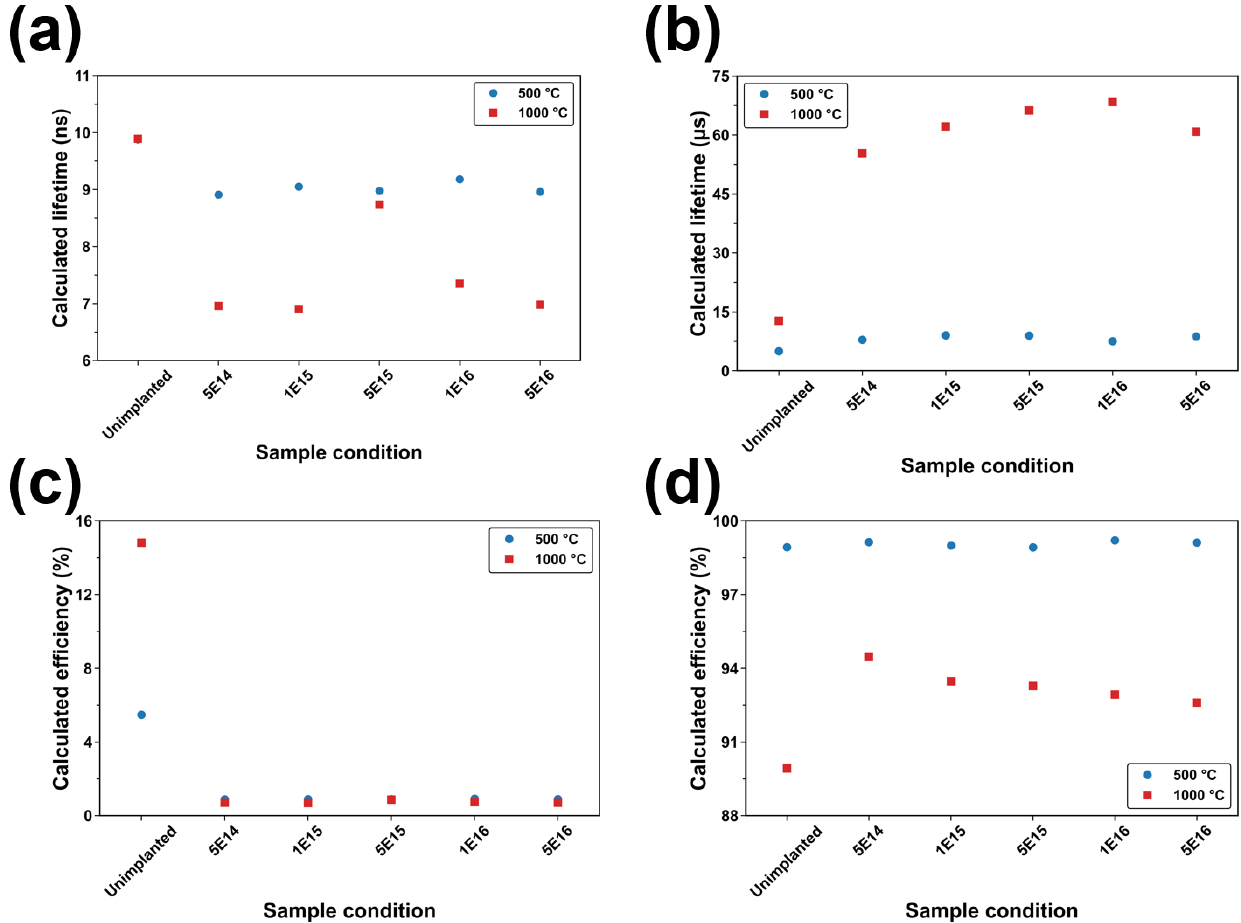}
      \caption{The radiative lifetime values obtained using the rate equation model for the (UVB+BB) PL, $\tau_1$, (a) and YB PL, $\tau_2$, (b). The PL efficiencies of the (UVB+BB) PL (c) and the YB PL (d).}
      \label{figure_REM_overall}
    \end{figure} 

Figure \ref{figure_REM_overall} shows the radiative lifetimes ($\tau_1, \tau_2$) and radiative efficiencies ($\eta_{UVB+BB}, \eta_{YB}$) obtained from the rate equation model as a function of implantation dose following annealing at 500 °C and 1000 °C. The radiative lifetime of the (UVB+BB) emission ($\tau_1$) exhibits a sensitivity to ion implantation dose and annealing temperature (Figure \ref{figure_REM_overall}(a)). The unimplanted control sits close to the 10 ns soft-penalty boundary enforced in the optimisation loss function. However, all implanted regions show a consistent downward shift into the range of 8.9-9.2 ns for the 500 °C series and further drop to 6.9-8.7 ns for the 1000 °C series. Crucially, because the optimiser incurs a mathematical penalty when pulling $\tau_1$ below 10 ns, this downward shift reflects a genuine, data-driven necessity to account for accelerated carrier loss via implantation-induced non-radiative pathways. 

In contrast, the YB PL lifetime ($\tau_2$) value extracted from the rate equation model demonstrates a strong annealing temperature dependency (Figure \ref{figure_REM_overall}(b)). At 500 °C, $\tau_2$ remains 4.9-9.0 \textmu s across the full range of ion doses. Following 1000 °C annealing, $\tau_2$ for all implanted doses is significantly increased to 55.3-68.4 \textmu s, signifying a severe reduction in the radiative recombination probability via this route. While we do not find a significant reduction in $N_{defect}$ (annihilation of deep level defects) or a boost in $k_{nr2}$ (proliferation of quenching centres), the increase in $\tau_2$ is likely to result from the partial passivation of the defect complexes responsible for the YB (evidenced by the marked decrease in $k_{r2}$).

Similarly, examining the radiative efficiencies obtained using the rate equation parameters further reveals the competitive balance between the two channels (Figures \ref{figure_REM_overall}(c) and \ref{figure_REM_overall}(d)). For the unimplanted control, $\eta_{\text{BB+UVB}}$ increases significantly from $5.48\%$ at 500 °C to $14.82\%$ at 1000 °C, reflecting thermal restoration of intrinsic material quality. However, across all implanted doses $\eta_{\text{BB+UVB}}$ remains capped at $< 1\%$ regardless of annealing temperature. This persistent suppression underscores that 1000 °C annealing is not sufficient to recover implantation-induced damage and may already cause catastrophic interdiffusion in the MQW structure (vide infra). For the YB channel, $\eta_{\text{YB}}$ maintains at $> 98.9\%$ across all areas at 500 °C. Elevating the annealing temperature to 1000 °C induces a systematic, uniform drop in $\eta_{\text{YB}}$ down to $89.9\%-94.5\%$. This reduction is a direct result of the decrease in the YB's radiative recombination rate ($k_{r2}$).

\subsection{Chromaticity coordinate analysis}
To evaluate the macroscopic colour-rendering behaviour and device-level functionality, CIE 1931 chromaticity coordinates were extracted for the 5E15 ions cm$^{-2}$ implanted area across all process stages and under varying excitation power densities (Figure \ref{figure_CIE_5E15}). Results for other In-implanted doses can be found in the Supplementary Information, section 4.

    \begin{figure}[htbp]
      \centering
      \includegraphics[width=1\textwidth]{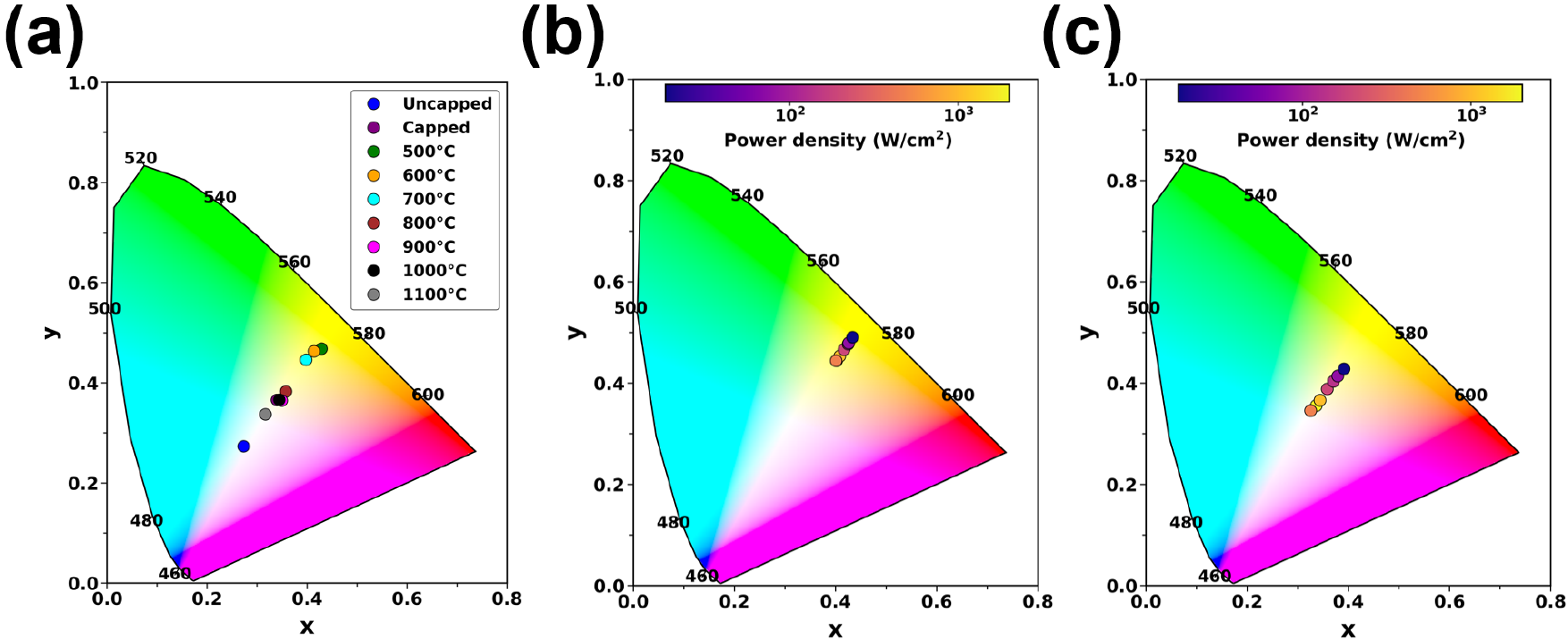}
      \caption{(a) CIE chromaticity diagrams obtained from the 5E15 ions cm$^{-2}$ implanted area using $\sim$2021 $\text{W/cm}^2$ 325 nm excitation as a function of annealing temperature. The variation as a function of excitation power density is shown following annealing at 500 °C (b) and 1000 °C (c).}
      \label{figure_CIE_5E15}
    \end{figure} 

The variation in global chromaticity across all thermal processing stages (Figure \ref{figure_CIE_5E15}(a)) systematically mirrors the microstructural defect evolution described in Figure \ref{Figure_intensity}. For the as-implanted state (uncapped/capped), the chromaticity coordinates reside in the cyan-blue region, corresponding to suppressed MQW emission with limited yellow background. Under the low/mid-temperature regime (500 – 700 °C), thermally activated YB-related defect complexes trigger a sharp shift toward the yellow/amber boundary. In this stage, YB radiative recombination overwhelms the MQW emission, dictating the dominant yellow emission. As the thermal budget increases further to the high-temperature regime (800 – 1100 °C), partial repair of lattice damage restores the MQW emission comparatively quicker than the YB emission. Consequently, the chromaticity coordinates retract back along a diagonal locus toward the achromatic white-light centre, demonstrating the feasibility of achieving balanced warm-to-cool white luminescence via controlled annealing.

Figures \ref{figure_CIE_5E15}(b) and \ref{figure_CIE_5E15}(c) respectively illustrate the excitation power density-dependent CIE coordinates under 500 °C and 1000 °C annealed conditions, highlighting a distinct contrast in colour rendering. Following 500 °C annealing, the chromaticity coordinates remain tightly pinned in the yellow-green zone across two orders of magnitude of excitation power density. Because 500 °C annealing fails to restore the MQW radiative pathway, carriers are predominantly captured by YB deep-level states regardless of photo-injection levels, yielding excitation-invariant yellow emission. Following 1000 °C annealing, in stark contrast, a continuous and linear chromaticity shift from the yellow/amber regime down to the cool white-blue zone is achieved upon increasing the power density. This enlarged colour-tuning trajectory provides direct macroscopic evidence of the differential recombination kinetics. Under low excitation power densities, unsaturated YB traps capture the majority of carriers; as the power density rises, the YB defect states saturate whilst the MQW pathway scales nearly linearly, shifting the spectral weight towards shorter wavelengths. 

The chromaticity coordinates ($x, y$) for all of the samples following each experimental stage, and as a function of excitation power density, are provided in the Supplementary Information, section 4. All implanted samples present very similar behaviour to that shown in Figure \ref{figure_CIE_5E15}. The behaviour of the reference unimplanted region has been fully described elsewhere\citep{liu2026thermal}.

The power density-dependent response further underscores the contrast between low- and high-temperature annealed states across all doses. At 500 °C, the severe suppression of the MQW pathway prevents effective carrier diversion, locking the macroscopic emission in the yellow/amber regime regardless of injection density. Varying the power density over two orders of magnitude produces negligible shifts in chromaticity for all implanted regions ($x \approx$ 0.40-0.43, $y \approx$ 0.44-0.49). Following 1000 °C annealing, an overall reduction in both x (from $\sim$0.39-0.40 down to $\sim$0.34-0.35) and y (from $\sim$0.43-0.44 down to $\sim$0.36-0.37) is observed as power density increases. This smooth chromaticity shift along a white-light trajectory occurs identically across all implanted doses. Physically, this universal tuning behaviour reflects the interplay between deep-trap saturation in the YB channel and nearly linear growth of the MQW emission. The dose-independent nature of this power density-tunable response demonstrates that ion-implantation-induced defect engineering offers a robust, highly reproducible strategy for excitation-driven colour control in monolithic GaN photonic platforms.

\subsection{Time-of-flight secondary ion mass spectrometry analysis}

\begin{figure}[htbp]
      \centering
      \includegraphics[width=1\textwidth]{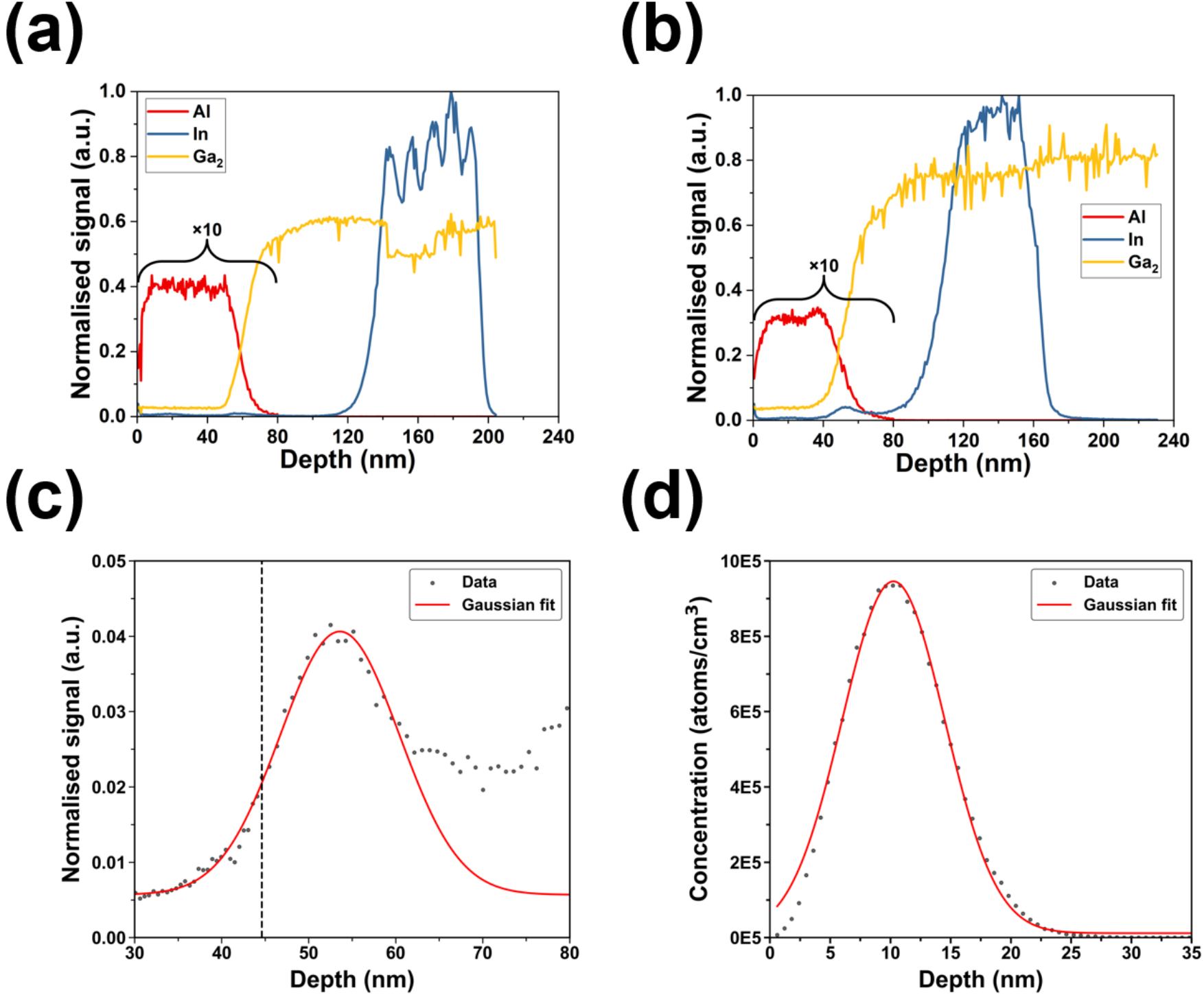}
      \caption{The ToF-SIMS profiles of (a) the unimplanted area and (b) the 5E15 ions cm$^{-2}$ implanted area, obtained using an ion current of 8 pA and 10 pA respectively. The Al signal in (a) and (b) has been multiplied by a factor of 10 for clarity. Gaussian fitting results of (c) the indium implant shoulder signal of the 5E15 ions cm$^{-2}$ implanted area (with a determined AlN/GaN interface marked by a black dashed line), and (d) the projected indium distribution following a SRIM simulation of 25 keV indium being implanted into a GaN matrix.}
      \label{figure_ToF_SIMS}
    \end{figure}

To determine the spatial distribution of the implanted indium and its influence on the embedded MQWs, ToF-SIMS depth profiling was performed after the 1100 °C annealing. Figures \ref{figure_ToF_SIMS}(a) and \ref{figure_ToF_SIMS}(b) present the element depth profiles of Al, In, and Ga between the unimplanted control and a 5E15 ions cm$^{-2}$ implanted area respectively. To prevent detector saturation, the gallium dimer (Ga$_2$) signal was selected for plotting because this species is less prevalent. Utilising the AlN and GaN sputter rates (4.59 nm and 7.76 nm per 1E14 ions/cm$^2$, respectively\citep{liu2026thermal}), the ToF-SIMS profiles are plotted using a depth scale. For the 5E15 ions cm$^{-2}$ implanted area, the calculated cumulative depth ($\sim$230.50 nm) based on these obtained AlN and GaN sputter rates matches well with the measured crater depth ($\sim$229.0 ±5.20 nm) obtained using AFM (see Supplementary Information, section 5).

For the unimplanted region, the indium profile exhibits a well-defined signal corresponding exclusively to the periodic MQW layer structure which remains intact following annealing as previously reported\citep{liu2026thermal}. By contrast, it is seen that implantation and annealing has significantly degraded the MQW structure. Furthermore, the presence of a weak shoulder signal preceding the main MQW signal is observed, attributed to the implanted 25 keV indium ions. Moreover, migration of the MQW signal towards the epitaxial GaN surface can be observed, which is considered to be collectively promoted by implantation-induced damage and post-implant thermal annealing\citep{yamaguchi2022atomic,yan2024influence}.

The ToF-SIMS results obtained from other dose-implanted areas are provided in Supplementary Information, section 5. While the macroscopic optical properties in Figures \ref{figure_mapping} and \ref{Figure_intensity} imply an implantation dose saturation occurring at or below the lowest dose used, the ToF-SIMS distributions reveal nuanced, dose-dependent diffusion kinetics. For a low dose of 5E14 ions cm$^{-2}$, an anomalous interdiffusion is observed, where the MQW profile severely broadens and completely submerges the shallow shoulder originating from ion-implanted In. For moderate doses from 1E15 to 1E16 ions cm$^{-2}$, the indium profiles maintain a relatively well-defined signal from where the MQW region resided, accompanied by a resolvable implanted In shoulder. For a high dose of 5E16 ions cm$^{-2}$, the indium profile exhibits a pronounced shift toward the sample surface, along with significant distortion of the MQW signal. 

It is speculated that at lower doses, isolated point defects dominate over extended defect clusters. Upon reaching 1100 °C, these mobile point defects mediate rapid In-Ga interdiffusion across the MQW boundaries prior to being annealed out. When increasing to moderate implantation doses, implantation-induced point defects partially coalesce into larger, less mobile defect clusters during initial heating, restricting catastrophic mass transport. At higher doses, the lattice damage exceeds the amorphisation threshold of GaN (>5E15 ions cm$^{-2}$,\citep{hernandez2017inxga1}). During ultra-high-temperature annealing, severe lattice dissociation, void formation, and phase decomposition drive substantial indium out-diffusion and surface segregation, shifting the apparent MQWs signal toward the capping layer.

Figure \ref{figure_ToF_SIMS}(c) presents the Gaussian fitting of the shoulder signal in Figure \ref{figure_ToF_SIMS}(b), yielding a peak position at 53.58 ±0.11 nm and a full width at half maximum (FWHM) of 16.14 ±0.36 nm. The uncertainty here and afterwards reflects the fit-covariance-based error. The obtained peak position itself is an indicator of the implant depth from the AlN/GaN interface. To account for the surface transient effect resulting in signal intermixing between elements, the AlN/GaN interface is determined as the critical point where the deeper-side Al signal intensity begins to fall below 75\% of its global plateau maximum. As such, the AlN/GaN interface is estimated to be located at $\sim$45 nm (marked by a black dashed line in Figure \ref{figure_ToF_SIMS}(c)), indicating an implant depth of $\sim$9 nm.

Figure \ref{figure_ToF_SIMS}(d) reveals the Gaussian fitting result of the projected indium distribution following a simulation of 25 keV indium being implanted into a GaN matrix via the Stopping and Range of Ions in Matter (SRIM) program\citep{ziegler1985stopping,williams1998ion}. It is found that the post-anneal indium shoulder peak has a considerably greater FWHM than the simulated post-implant counterpart (10.02 ±0.48 nm). The substantial extension of the FWHM from 10.02 nm to 16.14 nm may unveil the thermal diffusion and redistribution of the implanted indium during the multiple annealing processes. Meanwhile, the SRIM-simulated peak is centred at $\sim$10 nm. The slightly shallower shift towards 9 nm after annealing can be assigned to thermally induced outward migration.

\section{Conclusion}
In summary, we have comprehensively investigated the defect evolution kinetics, carrier recombination pathways, and macroscopic chromaticity engineering of indium-implanted InGaN/GaN MQWs as a function of In implantation dose and thermal processing up to 1100 °C. Spectral characterisation across process stages demonstrates that implantation-induced optical damage reaches complete saturation at an ion dose of 5E14 ions cm$^{-2}$. Thermal annealing at 500-700 °C activates deep-level states that dominate carrier recombination via YB emission, whereas high-temperature treatment (800-1100 °C) partially repairs lattice damage and restores excitonic BB emission transitions. Global optimisation using a two-channel rate-equation model reveals the quantitative mechanisms governing channel competition. Genuine implantation-induced non-radiative losses are reflected by the downward shift of the BB PL lifetime below the optimisation penalty boundary (from $\sim$10 to 6.9-9.2 ns) and a significant efficiency drop to below 1\%. 

Meanwhile, increasing the annealing temperature from 500 °C to 1000 °C causes the implanted areas' YB emission calculated lifetime to increase (from 7.5-9.0 \textmu s to 55.3-68.4 \textmu s) while slightly reducing the efficiency ($>$98.9$\%$ to 92.6-94.5$\%$), predominantly due to the passivation of deep level states. Mapped onto CIE chromaticity space, the 1000 °C annealed post-implant areas exhibit a highly reproducible, excitation power density tunable colour shift along a warm-yellow to cool-white/blue trajectory ($x: 0.39-0.40 \rightarrow 0.34-0.35, y: 0.43-0.44 \rightarrow 0.36-0.37$). The universal overlap of this chromaticity trajectory across all doses highlights the robustness of ion-implantation-based defect engineering for excitation-controlled monolithic optoelectronic devices.

\section{Acknowledgements}

This work was funded by EPSRC grant EP/V001914/1, and by capital investment by the University of Manchester. Q-S.L thanks the China Scholarship Council for financial support. The authors thank R Oliver (University of Cambridge) for providing the sample studied in this work, and D Binks (University of Manchester) and S Church (University of Salford) for useful discussions.

\section{Author contributions statement}

Q-S.L, M.A, S.S, M.C and S.C conducted all experimental work. Q-S.L, S.S, M.C and R.C performed data analysis. N.L, D.B and R.C supervised the work. All authors contributed to the writing of the manuscript.

\section{Competing interests}

The author(s) declare no competing interests.

\bibliographystyle{unsrtnat}
\bibliography{references}  

\clearpage
\begin{center}
    \Large
    \textbf{Supplementary information}
    \\[20pt]
    \normalsize 
\end{center}
\setcounter{figure}{0}
\setcounter{table}{0}
\setcounter{section}{0}
\setcounter{equation}{0}
\renewcommand*{\thefigure}{S\arabic{figure}}
\renewcommand*{\thetable}{S\arabic{table}}
\renewcommand*{\theequation}{S\arabic{equation}}

\section{Steady-state photoluminescence spectrum analysis for other implanted areas}
\label{SectionS1}
\begin{figure}[htbp]
      \centering
      \includegraphics[width=1\textwidth]{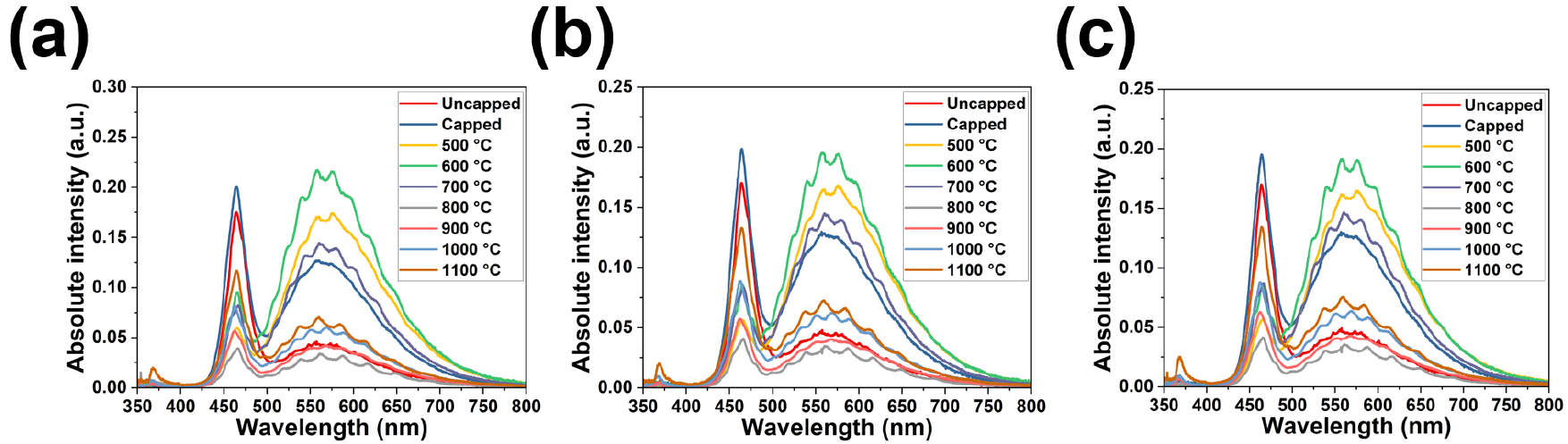}
      \caption{Room temperature PL spectra at all stages of processing for regions implanted with In doses of 1E15 ions cm$^{-2}$ (a), 5E15 ions cm$^{-2}$ (b), and 1E16 ions cm$^{-2}$ (c).}
      \label{figureS_RT_PL_1E15_5E15_1E16}
    \end{figure}

The spectral evolution across progressive thermal annealing stages is detailed in Figures \ref{figureS_RT_PL_1E15_5E15_1E16}(a) to \ref{figureS_RT_PL_1E15_5E15_1E16}(c) for doping concentrations of 1E15 ions cm$^{-2}$, 5E15 ions cm$^{-2}$, and 5E16 ions cm$^{-2}$ respectively.

\section{Power-dependent measurements for other implanted areas}
\label{SectionS2}
\begin{figure}[htbp]
      \centering
      \includegraphics[width=1\textwidth]{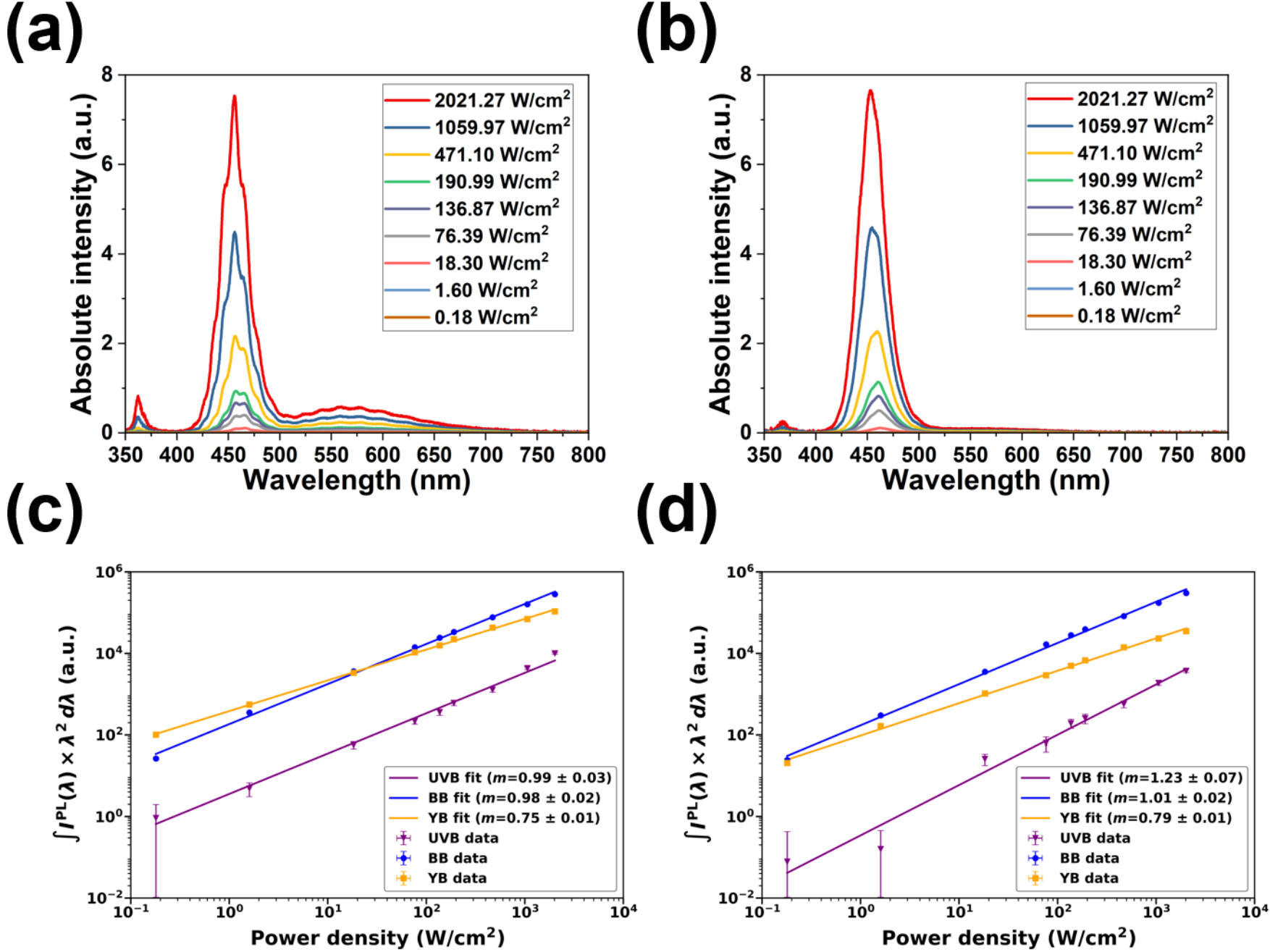}
      \caption{For the unimplanted area. Excitation power density dependence of 325 nm excited PL obtained after (a) 500 °C and (b) 1000 °C annealing. Log-Log plot of the integrated PL intensity as a function of excitation power density for the main emission peaks obtained after (c) 500 °C and (d) 1000 °C annealing.}
      \label{figureS_PD_PL_unimplanted}
    \end{figure}

\begin{figure}[htbp]
      \centering
      \includegraphics[width=1\textwidth]{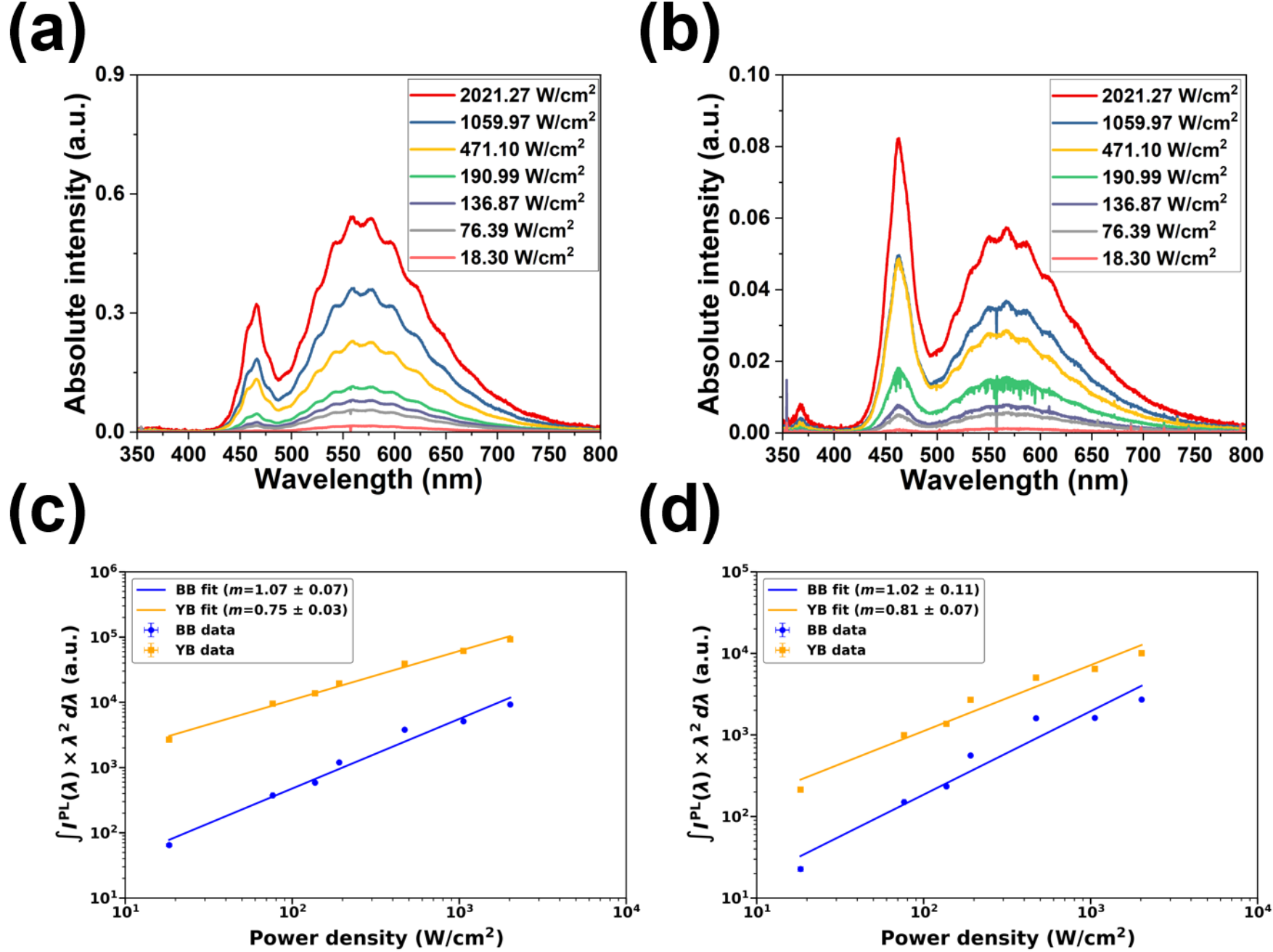}
      \caption{For the 5E14 ions cm$^{-2}$ In implanted area. Excitation power density dependence of 325 nm excited PL obtained after (a) 500 °C and (b) 1000 °C annealing. Log-Log plot of the integrated PL intensity as a function of excitation power density for the main emission peaks obtained after (c) 500 °C and (d) 1000 °C annealing.}
      \label{figureS_PD_PL_5E14}
    \end{figure}

\begin{figure}[htbp]
      \centering
      \includegraphics[width=1\textwidth]{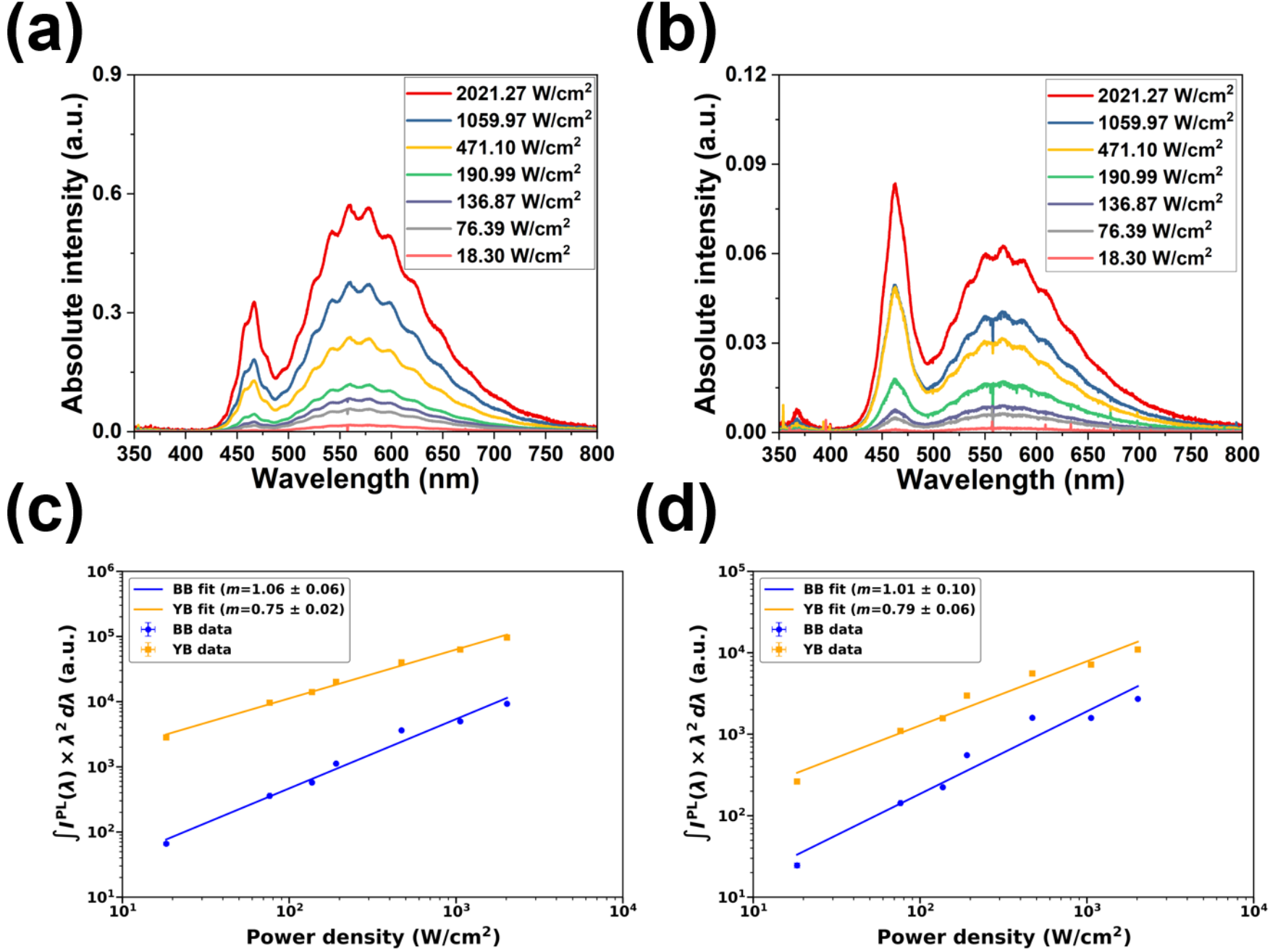}
      \caption{For the 1E15 ions cm$^{-2}$ In implanted area. Excitation power density dependence of 325 nm excited PL obtained after (a) 500 °C and (b) 1000 °C annealing. Log-Log plot of the integrated PL intensity as a function of excitation power density for the main emission peaks obtained after (c) 500 °C and (d) 1000 °C annealing.}
      \label{figureS_PD_PL_1E15}
    \end{figure}

\begin{figure}[htbp]
      \centering
      \includegraphics[width=1\textwidth]{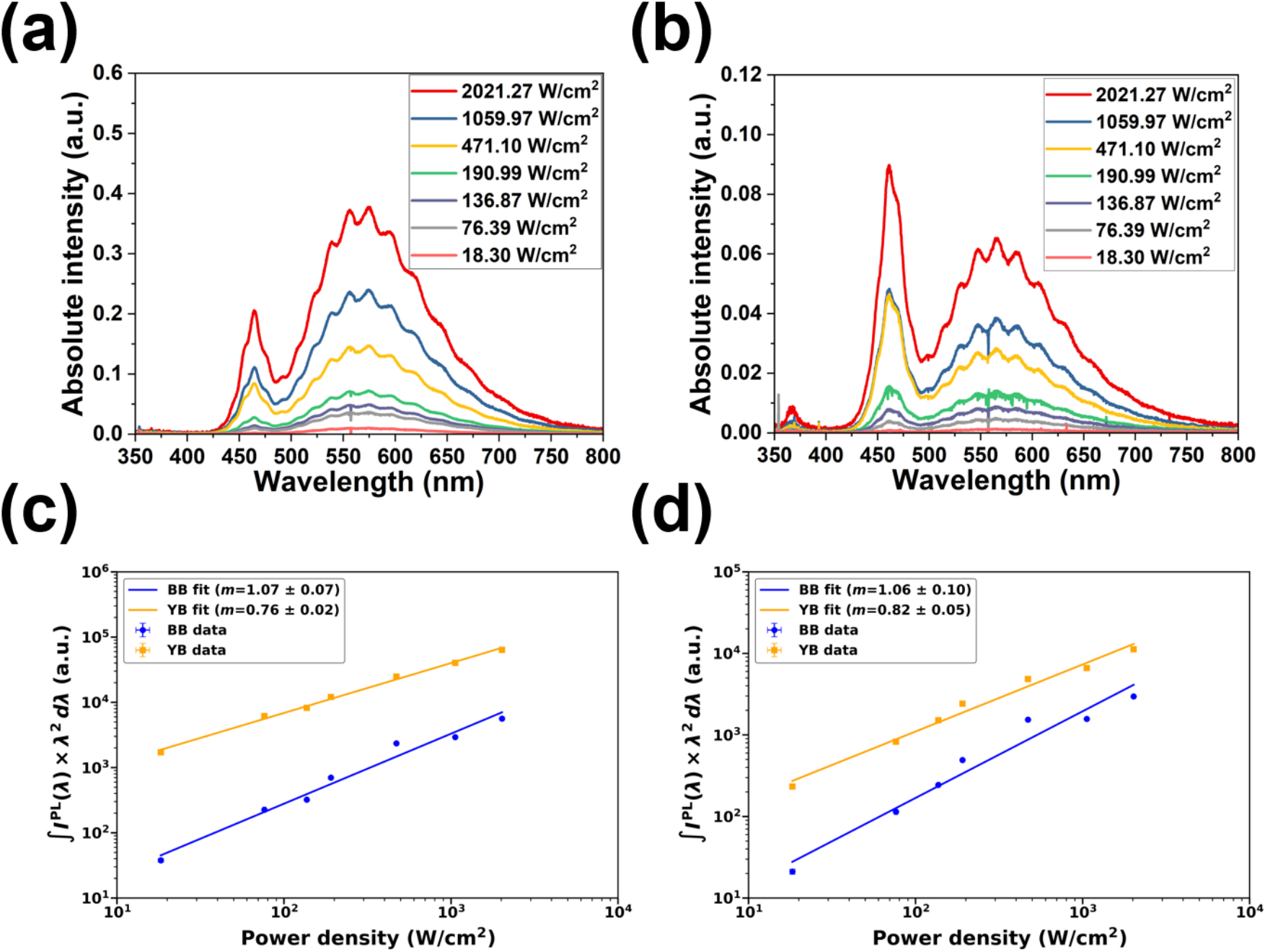}
      \caption{For the 1E16 ions cm$^{-2}$ In implanted area. Excitation power density dependence of 325 nm excited PL obtained after (a) 500 °C and (b) 1000 °C annealing. Log-Log plot of the integrated PL intensity as a function of excitation power density for the main emission peaks obtained after (c) 500 °C and (d) 1000 °C annealing.}
      \label{figureS_PD_PL_1E16}
    \end{figure}

\begin{figure}[htbp]
      \centering
      \includegraphics[width=1\textwidth]{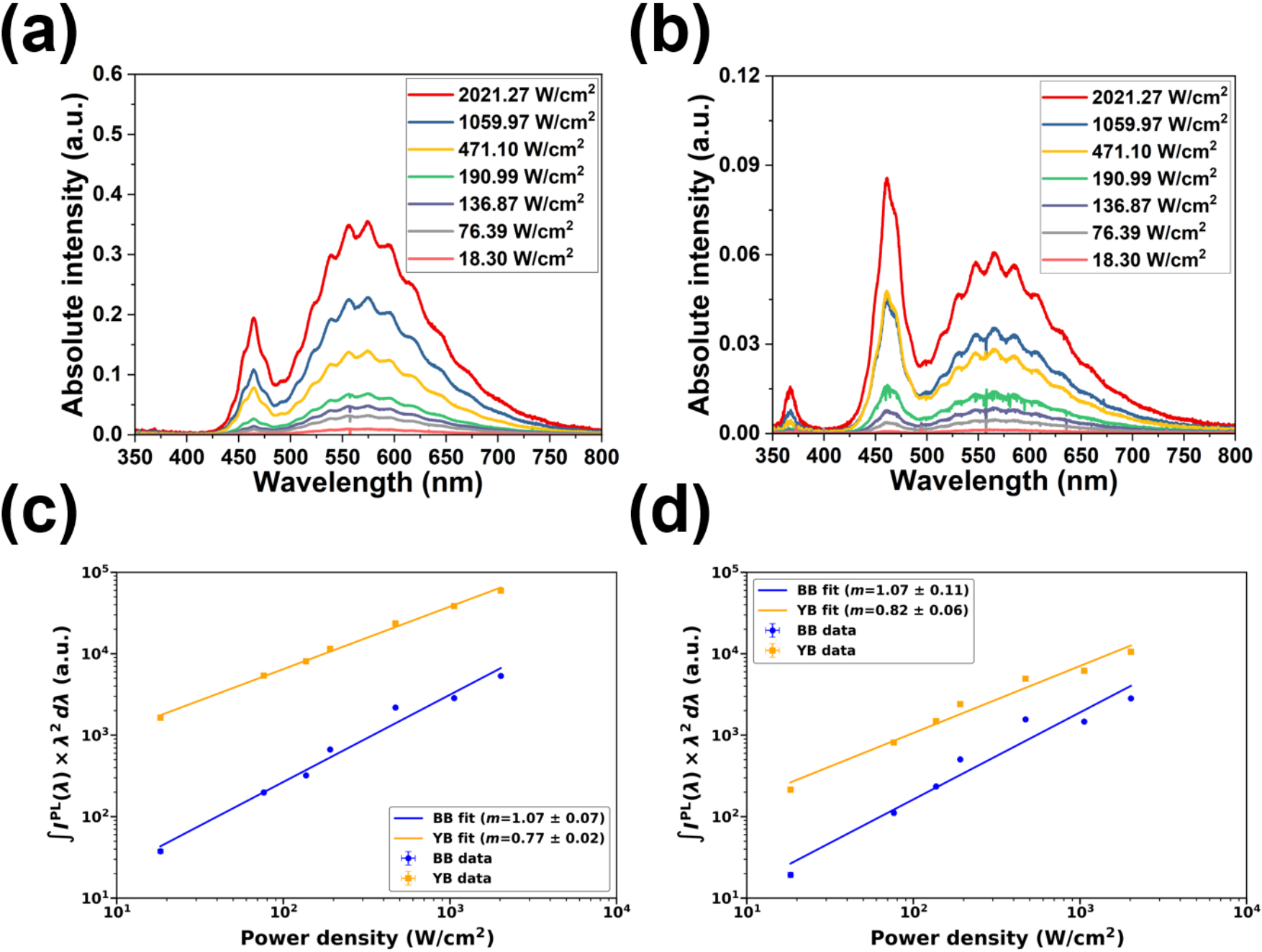}
      \caption{For the 5E16 ions cm$^{-2}$ In implanted area. Excitation power density dependence of 325 nm excited PL obtained after (a) 500 °C and (b) 1000 °C annealing. Log-Log plot of the integrated PL intensity as a function of excitation power density for the main emission peaks obtained after (c) 500 °C and (d) 1000 °C annealing.}
      \label{figureS_PD_PL_5E16}
    \end{figure}

\begin{figure}[htbp]
      \centering
      \includegraphics[width=1\textwidth]{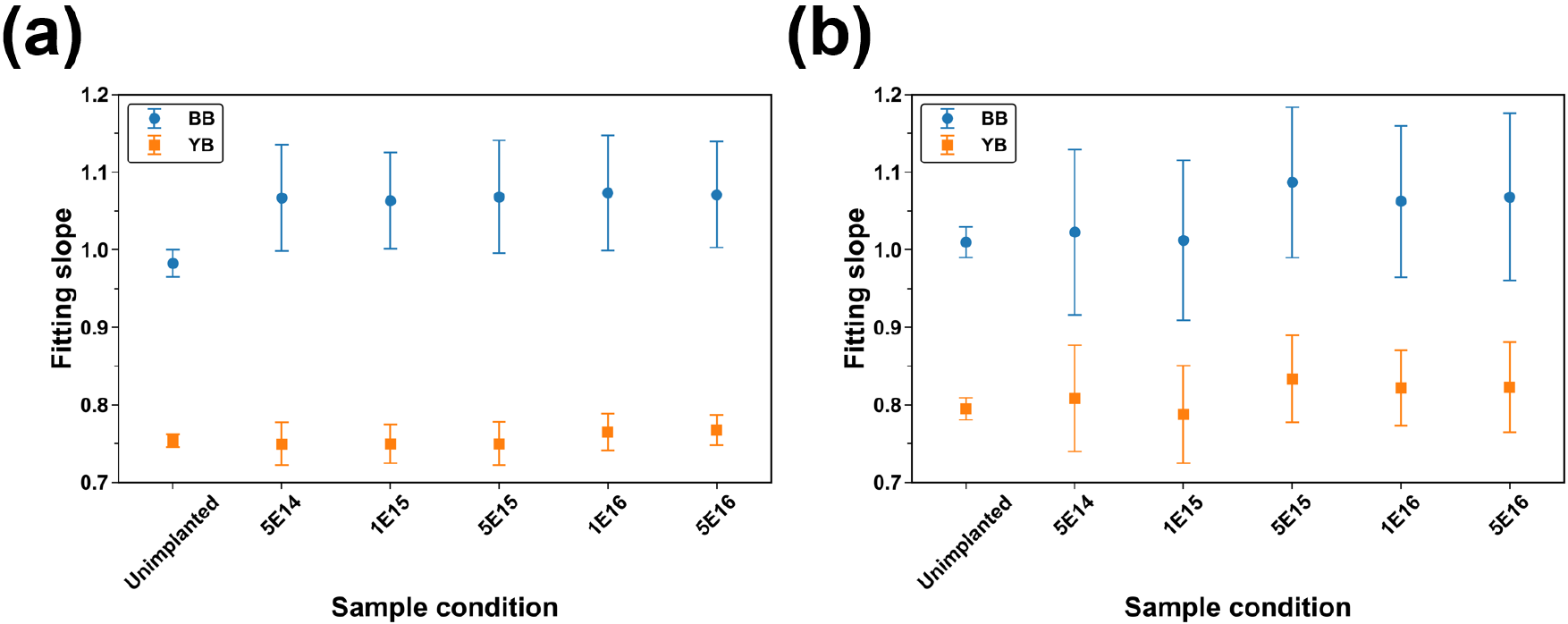}
      \caption{Power-law fitted slope of the excitation power density BB and YB PL intensity for the unimplanted and implanted areas after annealing at (a) 500 °C and (b) 1000 °C.}
      \label{figureS_slope}
    \end{figure}

Figures \ref{figureS_PD_PL_unimplanted} to \ref{figureS_PD_PL_5E16} respectively depict the excitation power density-dependent PL spectra and power-law slope analysis of the unimplanted, the 5E14 ions cm$^{-2}$, the 1E15 ions cm$^{-2}$, the 1E16 ions cm$^{-2}$, and the 5E16 ions cm$^{-2}$ In implanted areas following 500 °C and 1000 °C annealing. Note that the results of the unimplanted reference contain two additional data points measured under 1.6 $\text{W/cm}^2$ and 0.18 $\text{W/cm}^2$. This is because these data points still show distinguishable emission from background noise, unlike the implanted counterparts. Figure \ref{figureS_slope} shows the power-law exponents ($m$) for the BB and YB PL excitation power density dependence.

\section{Rate equation modelling of implanted areas}
\label{SectionS3}
\begin{figure}[htbp]
      \centering
      \includegraphics[width=1\textwidth]{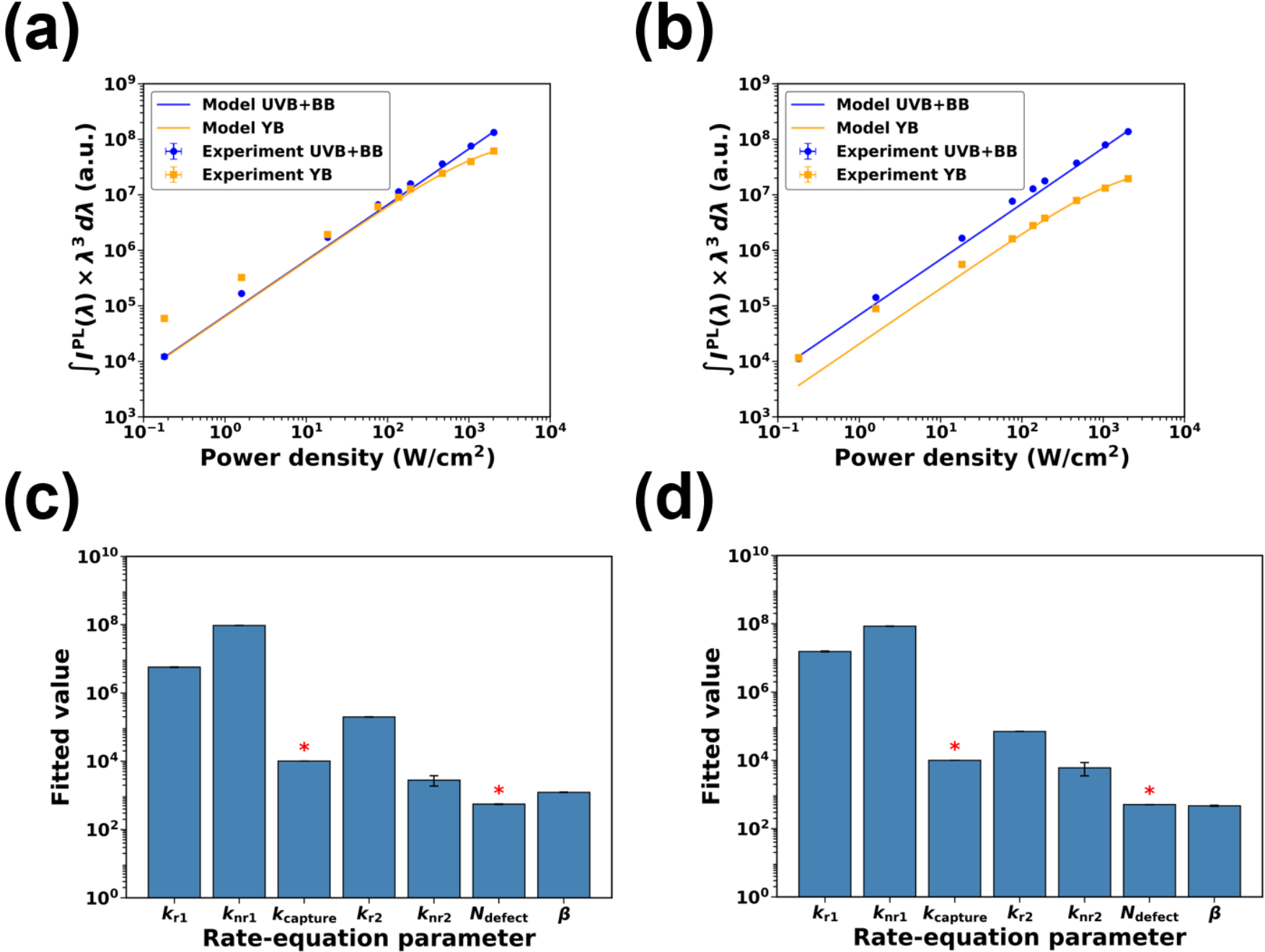}
      \caption{For the unimplanted area. Comparison between the emission channels' experimental PL integrated intensities (data points) and the best fit obtained using the rate-equation model (lines) as a function of excitation power density after annealing at (a) 500 °C and (b) 1000 °C. Mean fitting parameter values and standard deviations obtained from repeated differential evolution runs after annealing at (c) 500 °C and (d) 1000 °C. An `*' identifies a parameter whose best-fit value lies within 5\% of a search-space boundary.}
      \label{figureS_REM_unimplanted}
    \end{figure}

\begin{figure}[htbp]
      \centering
      \includegraphics[width=1\textwidth]{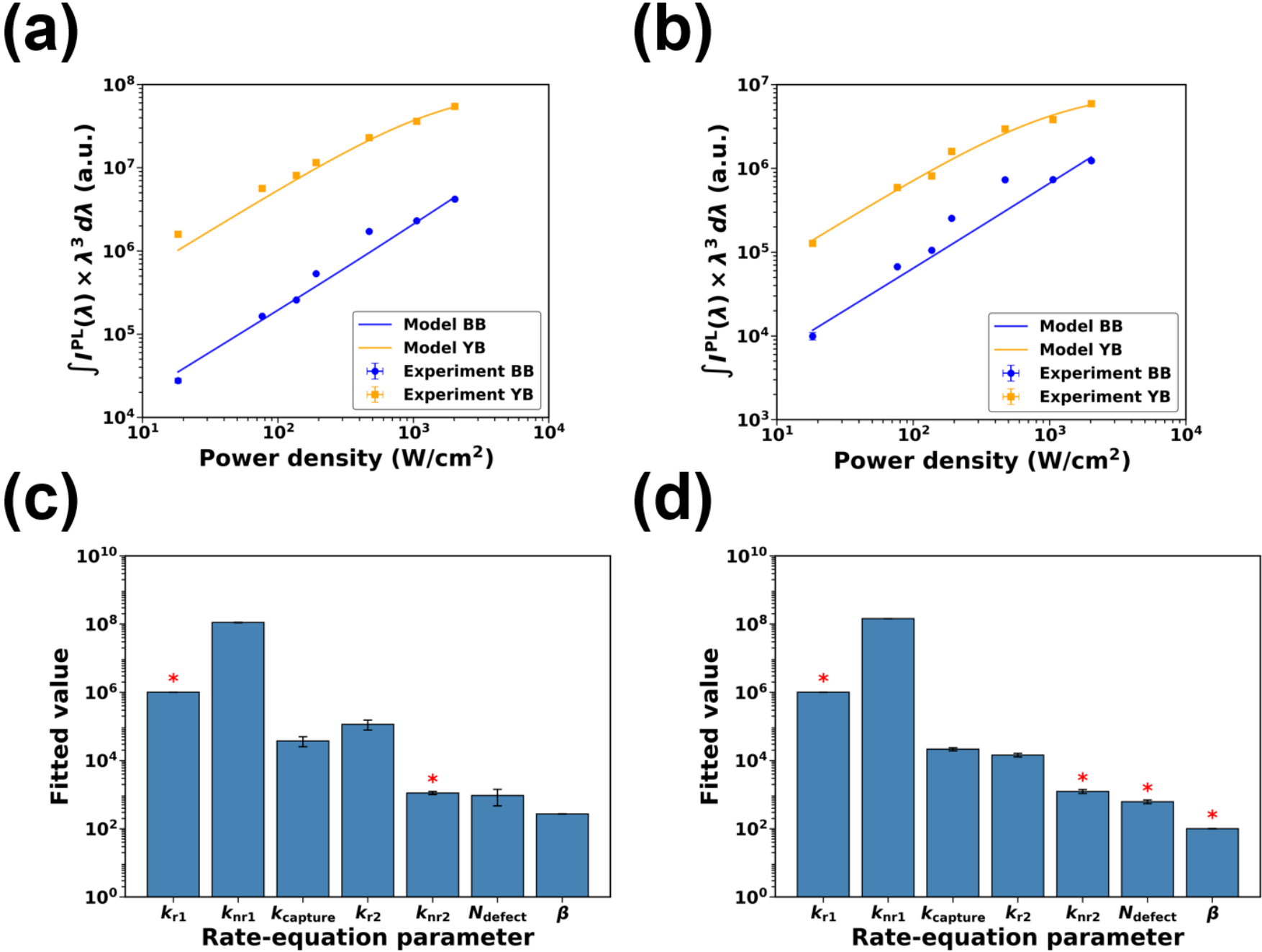}
      \caption{For the 5E14 ions cm$^{-2}$ In implanted area. Comparison between the emission channels' experimental PL integrated intensities (data points) and the best fit obtained using the rate-equation model (lines) as a function of excitation power density after annealing at (a) 500 °C and (b) 1000 °C. Mean fitting parameter values and standard deviations obtained from repeated differential evolution runs after annealing at (c) 500 °C and (d) 1000 °C. An `*' identifies a parameter whose best-fit value lies within 5\% of a search-space boundary.}
      \label{figureS_REM_5E14}
    \end{figure}

\begin{figure}[htbp]
      \centering
      \includegraphics[width=1\textwidth]{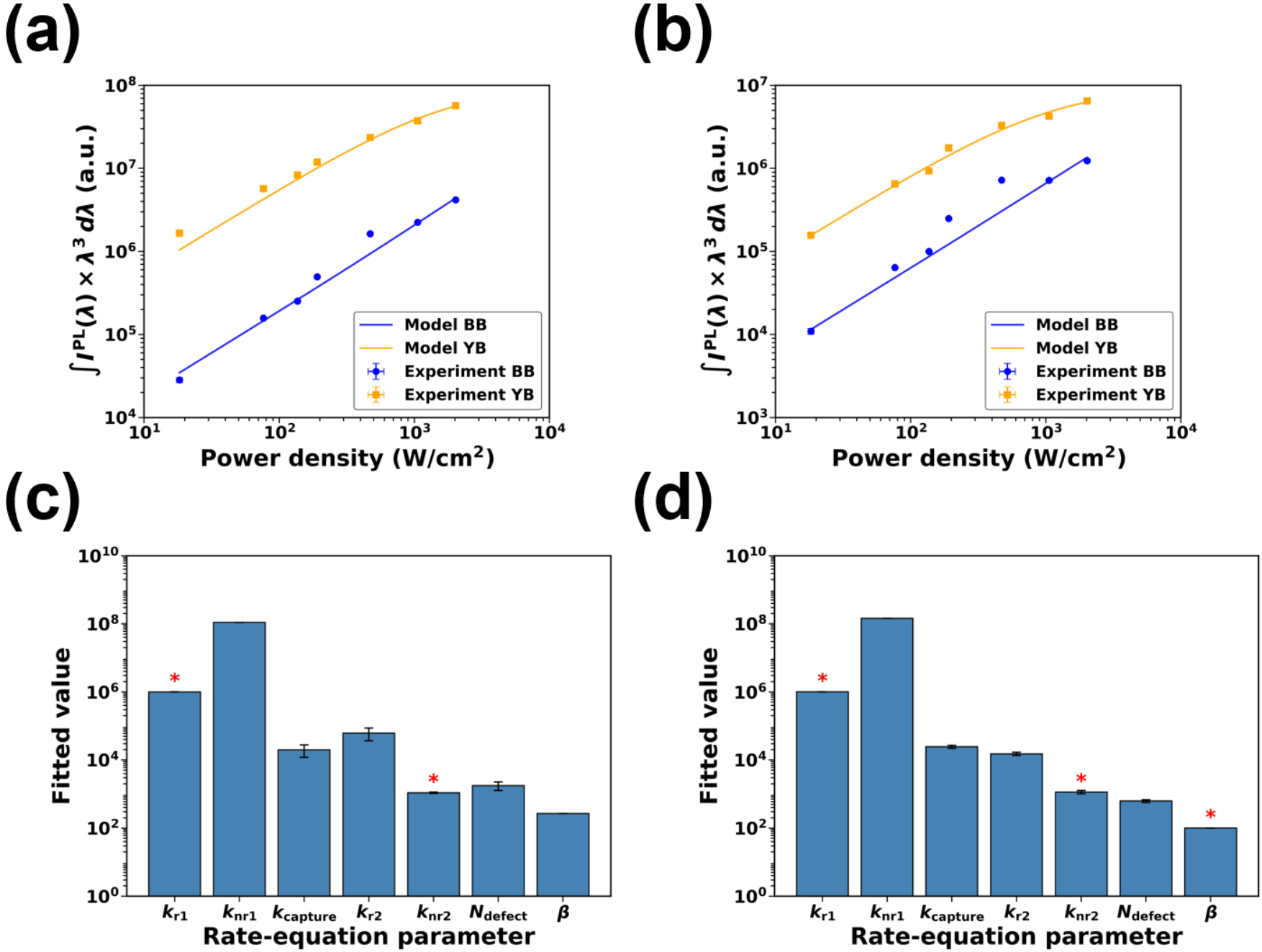}
      \caption{For the 1E15 ions cm$^{-2}$ In implanted area. Comparison between the emission channels' experimental PL integrated intensities (data points) and the best fit obtained using the rate-equation model (lines) as a function of excitation power density after annealing at (a) 500 °C and (b) 1000 °C. Mean fitting parameter values and standard deviations obtained from repeated differential evolution runs after annealing at (c) 500 °C and (d) 1000 °C. An `*' identifies a parameter whose best-fit value lies within 5\% of a search-space boundary.}
      \label{figureS_REM_1E15}
    \end{figure}

\begin{figure}[htbp]
      \centering
      \includegraphics[width=1\textwidth]{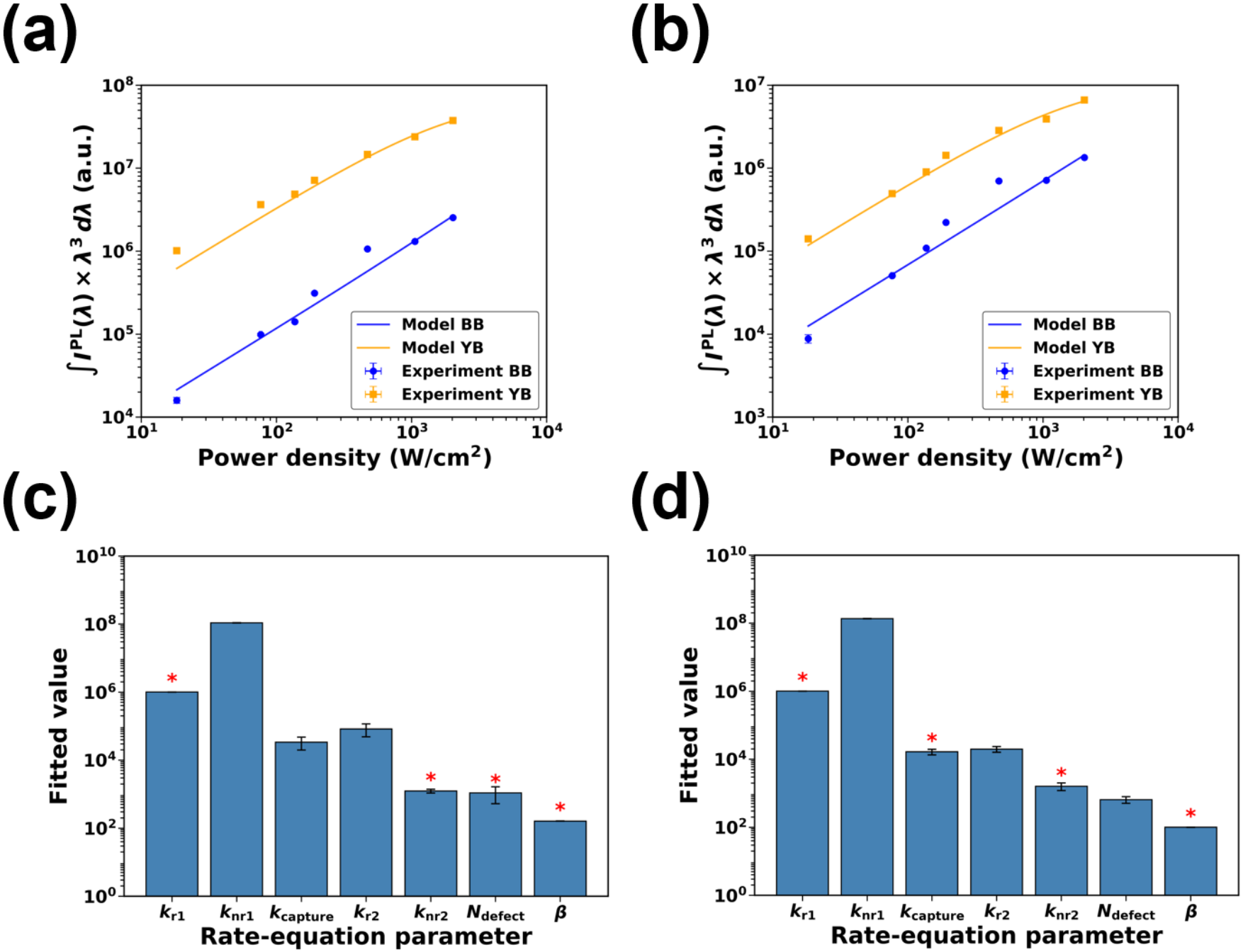}
      \caption{For the 1E16 ions cm$^{-2}$ In implanted area. Comparison between the emission channels' experimental PL integrated intensities (data points) and the best fit obtained using the rate-equation model (lines) as a function of excitation power density after annealing at (a) 500 °C and (b) 1000 °C. Mean fitting parameter values and standard deviations obtained from repeated differential evolution runs after annealing at (c) 500 °C and (d) 1000 °C. An `*' identifies a parameter whose best-fit value lies within 5\% of a search-space boundary.}
      \label{figureS_REM_1E16}
    \end{figure}

\begin{figure}[htbp]
      \centering
      \includegraphics[width=1\textwidth]{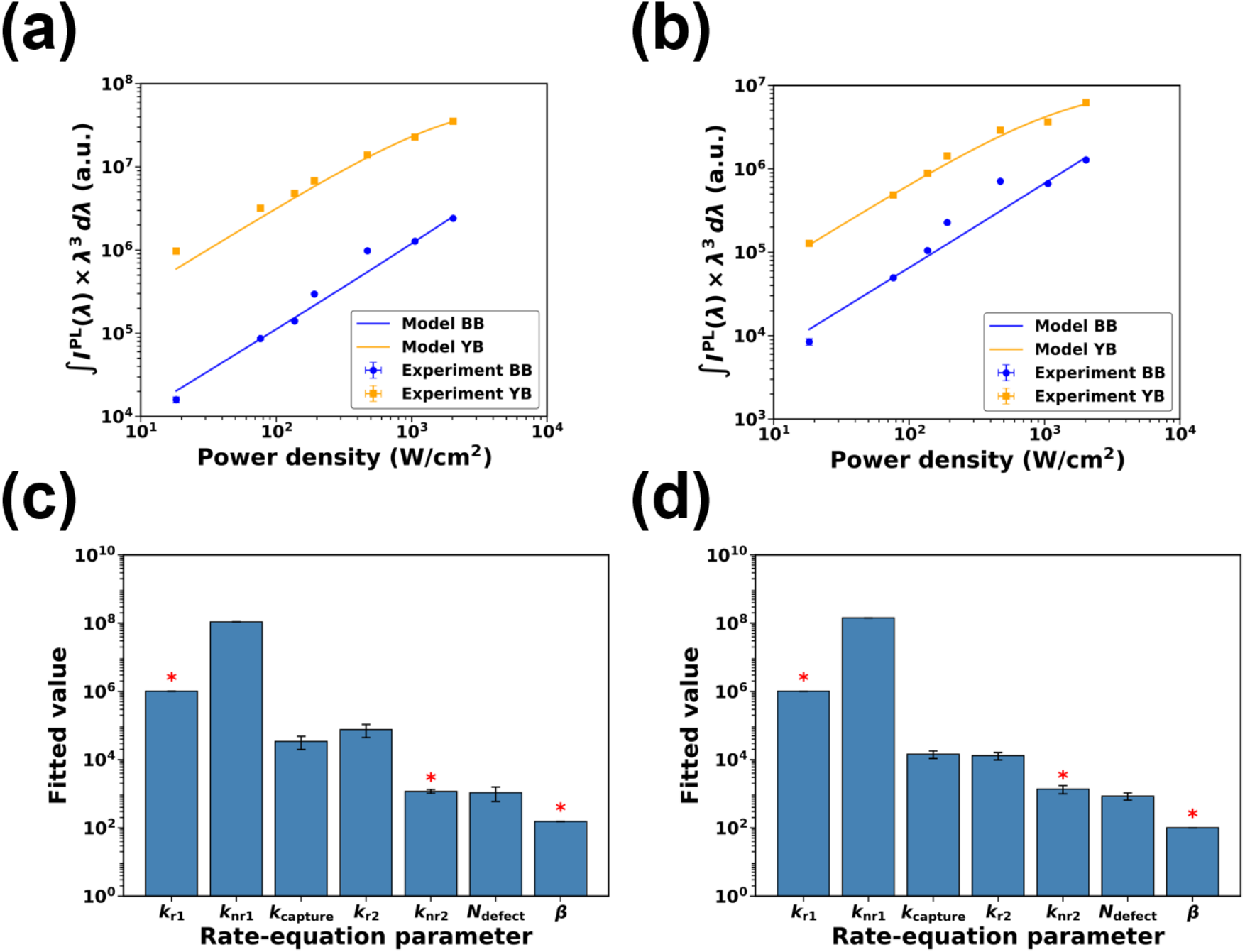}
      \caption{For the 5E16 ions cm$^{-2}$ In implanted area. Comparison between the emission channels' experimental PL integrated intensities (data points) and the best fit obtained using the rate-equation model (lines) as a function of excitation power density after annealing at (a) 500 °C and (b) 1000 °C. Mean fitting parameter values and standard deviations obtained from repeated differential evolution runs after annealing at (c) 500 °C and (d) 1000 °C. An `*' identifies a parameter whose best-fit value lies within 5\% of a search-space boundary.}
      \label{figureS_REM_5E16}
    \end{figure}

Figures \ref{figureS_REM_unimplanted} to \ref{figureS_REM_5E16} respectively present the rate equation modelling output results for the unimplanted, the 5E14 ions cm$^{-2}$, the 1E15 ions cm$^{-2}$, the 1E16 ions cm$^{-2}$, and the 5E16 ions cm$^{-2}$ In implanted areas following 500 °C and 1000 °C annealing.

\begin{figure}[htbp]
      \centering
      \includegraphics[width=1\textwidth]{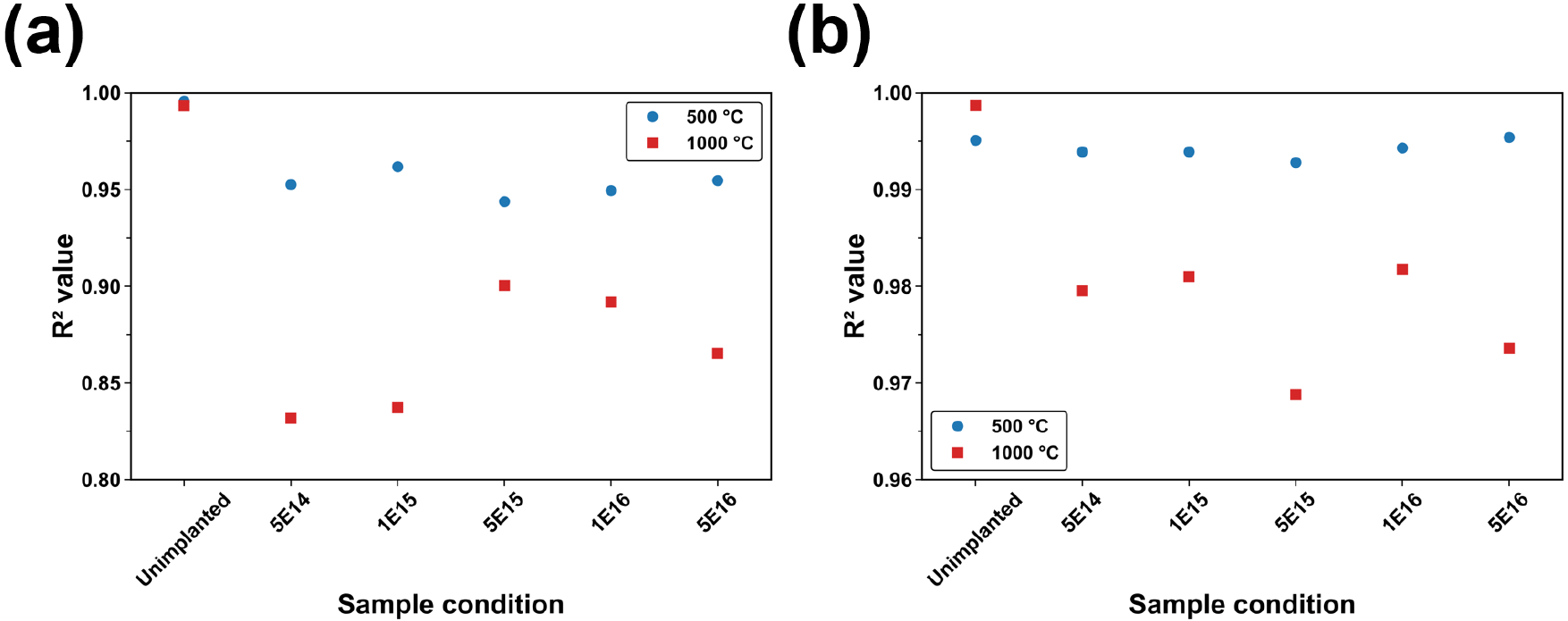}
      \caption{Across all fitting scenarios, the $R^2$ values of (a) the UVB plus BB channel and (b) the YB channel.}
      \label{figureS_R2_value}
    \end{figure}

In the nonlinear regression framework, the coefficient of determination ($R^2$), formulated in Equation \ref{equ:coefficient of determination}, quantifies the proportion of variance in the experimental data ($y_i^{\mathrm{Exp}}$) accounted for by the model predictions ($y_i^{\mathrm{Model}}$) relative to the experimental mean ($\overline{y}^{\mathrm{Exp}}$). Serving as a standard metric for goodness of fit, an $R^2$ value approaching unity indicates optimal predictive accuracy, whereas a value near zero implies explanatory performance equivalent to a horizontal baseline model. The distribution of these $R^2$ values across all evaluated fitting scenarios is detailed in Figure \ref{figureS_R2_value}.
    
    \begin{equation}
        R^2 = 1 - \frac{\sum\limits_{i=1}^{n}\left(y_i^{\mathrm{Exp}} - y_i^{\mathrm{Model}}\right)^2}{\sum\limits_{i=1}^{n}\left(y_i^{\mathrm{Exp}} - \overline{y}^{\mathrm{Exp}}\right)^2} \label{equ:coefficient of determination} 
    \end{equation}
    
\section{CIE 1931 chromaticity diagrams for other implanted areas}
\label{SectionS4}
\begin{figure}[htbp]
      \centering
      \includegraphics[width=0.8\textwidth]{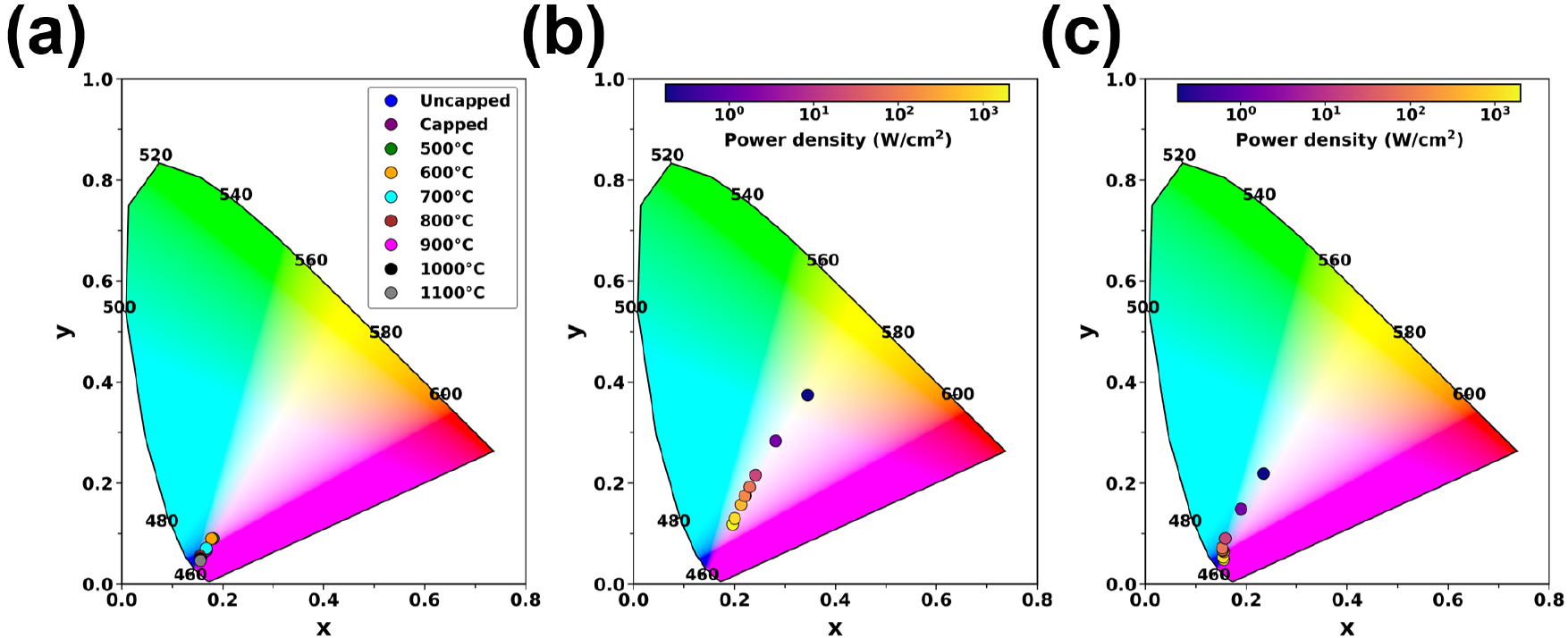}
      \caption{For the unimplanted area. CIE chromaticity diagrams obtained using (a) 2021.27 $\text{W/cm}^2$ 325 nm excitation as a function of annealing temperature, and as a function of excitation power density following annealing at (b) 500 °C and (c) 1000 °C.}
      \label{figureS_CIE_unimplanted}
    \end{figure}

\begin{figure}[htbp]
      \centering
      \includegraphics[width=0.8\textwidth]{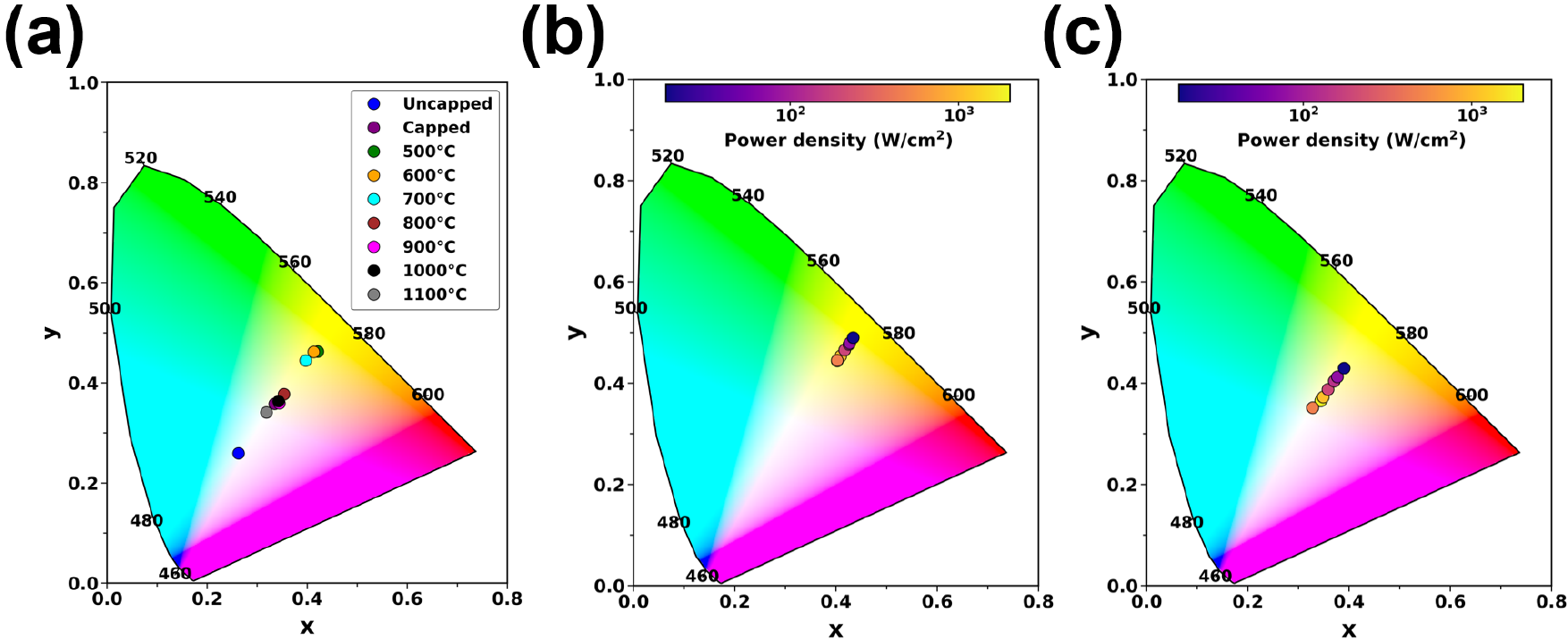}
      \caption{For the 5E14 ions cm$^{-2}$ In implanted area. CIE chromaticity diagrams obtained using (a) 2021.27 $\text{W/cm}^2$ 325 nm excitation as a function of annealing temperature, and as a function of excitation power density following annealing at (b) 500 °C and (c) 1000 °C.}
      \label{figureS_CIE_5E14}
    \end{figure}

\begin{figure}[htbp]
      \centering
      \includegraphics[width=0.8\textwidth]{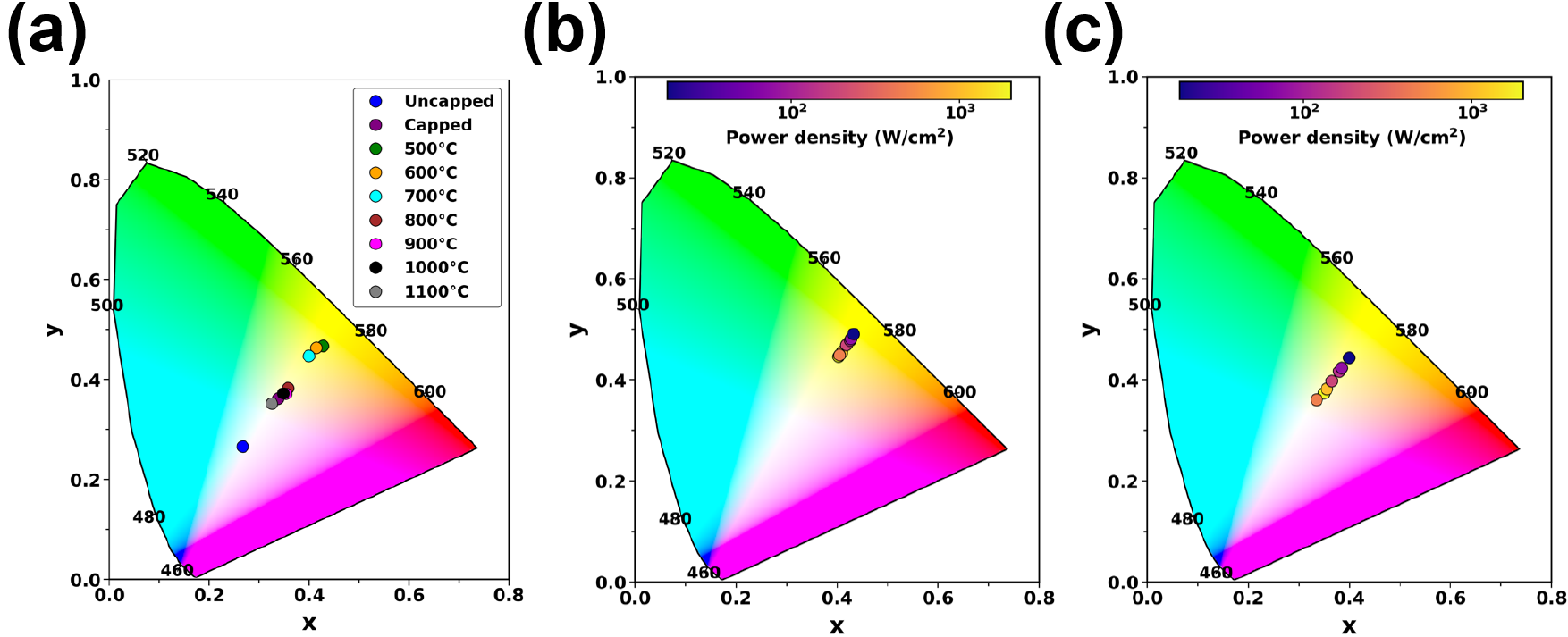}
      \caption{For the 1E15 ions cm$^{-2}$ In implanted area. CIE chromaticity diagrams obtained using (a) 2021.27 $\text{W/cm}^2$ 325 nm excitation as a function of annealing temperature, and as a function of excitation power density following annealing at (b) 500 °C and (c) 1000 °C.}
      \label{figureS_CIE_1E15}
    \end{figure}

\begin{figure}[htbp]
      \centering
      \includegraphics[width=0.8\textwidth]{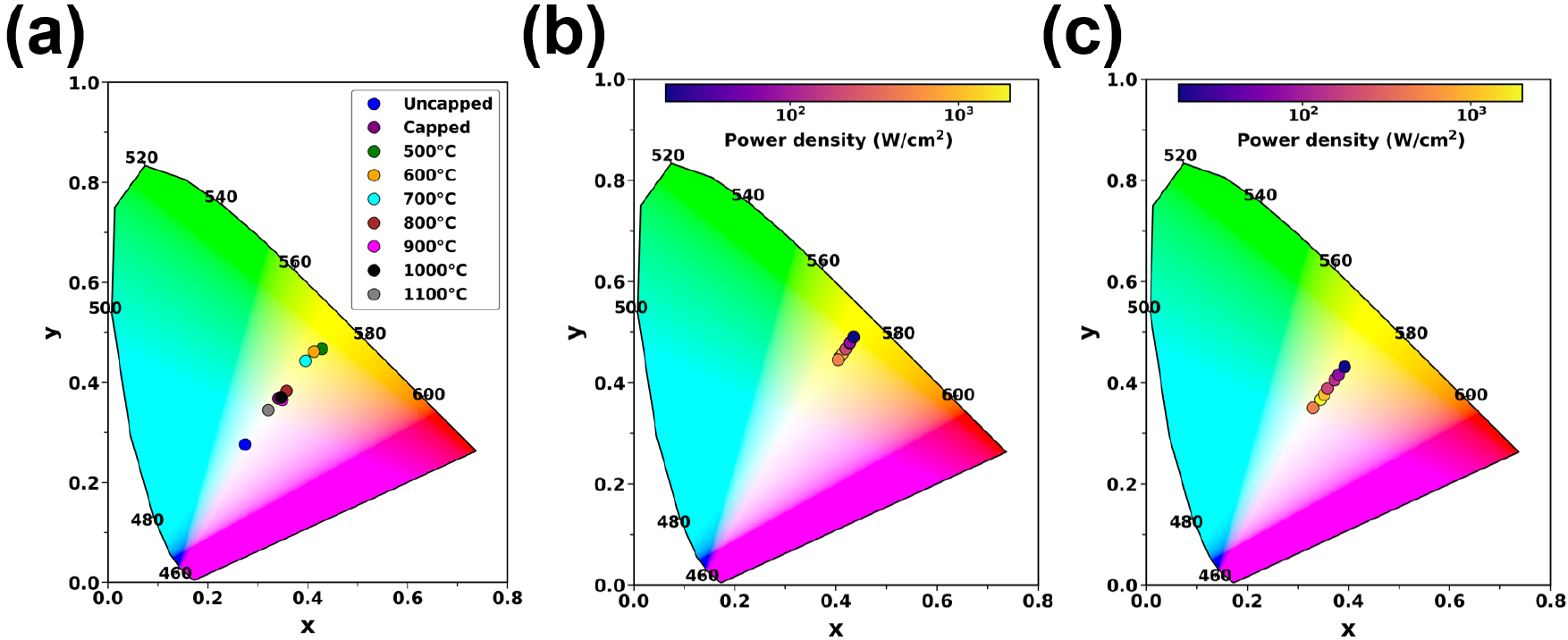}
      \caption{For the 1E16 ions cm$^{-2}$ In implanted area. CIE chromaticity diagrams obtained using (a) 2021.27 $\text{W/cm}^2$ 325 nm excitation as a function of annealing temperature, and as a function of excitation power density following annealing at (b) 500 °C and (c) 1000 °C.}
      \label{figureS_CIE_1E16}
    \end{figure}

\begin{figure}[htbp]
      \centering
      \includegraphics[width=0.8\textwidth]{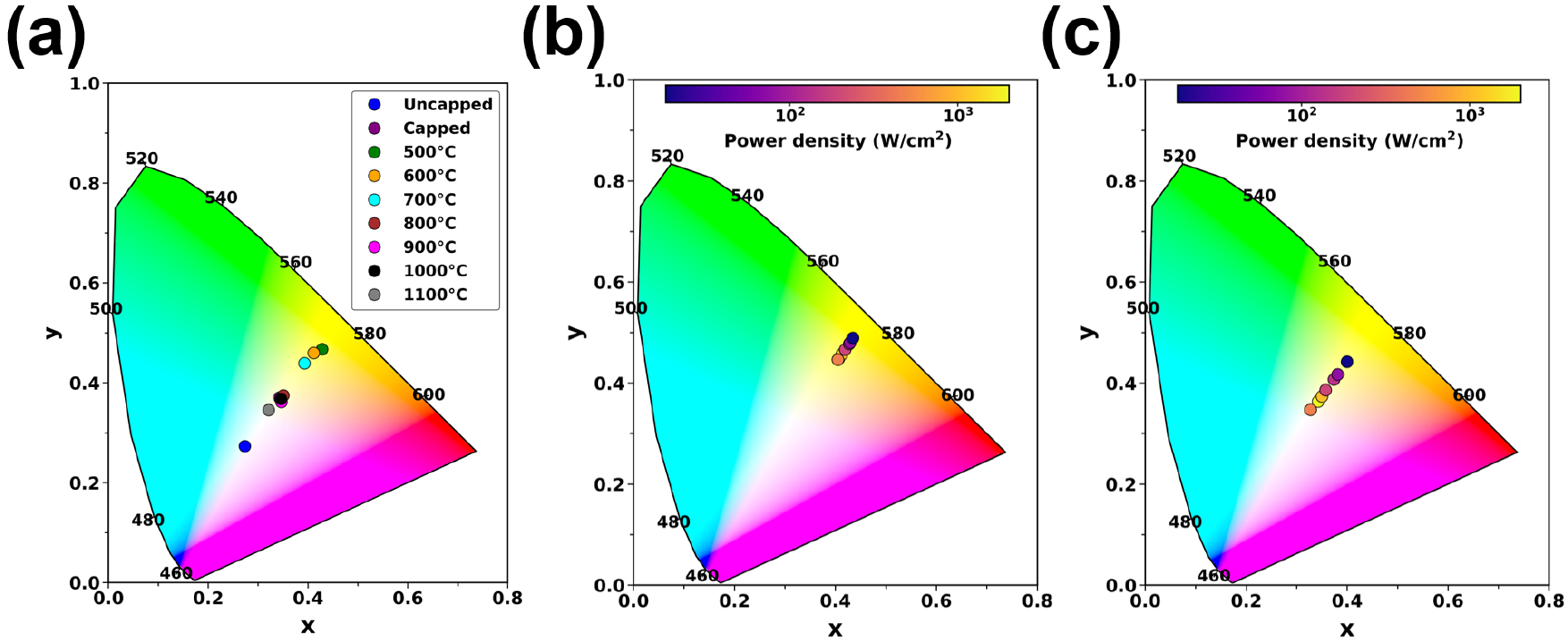}
      \caption{For the 5E16 ions cm$^{-2}$ In implanted area. CIE chromaticity diagrams obtained using (a) 2021.27 $\text{W/cm}^2$ 325 nm excitation as a function of annealing temperature, and as a function of excitation power density following annealing at (b) 500 °C and (c) 1000 °C.}
      \label{figureS_CIE_5E16}
    \end{figure}

\begin{figure}[htbp]
      \centering
      \includegraphics[width=1\textwidth]{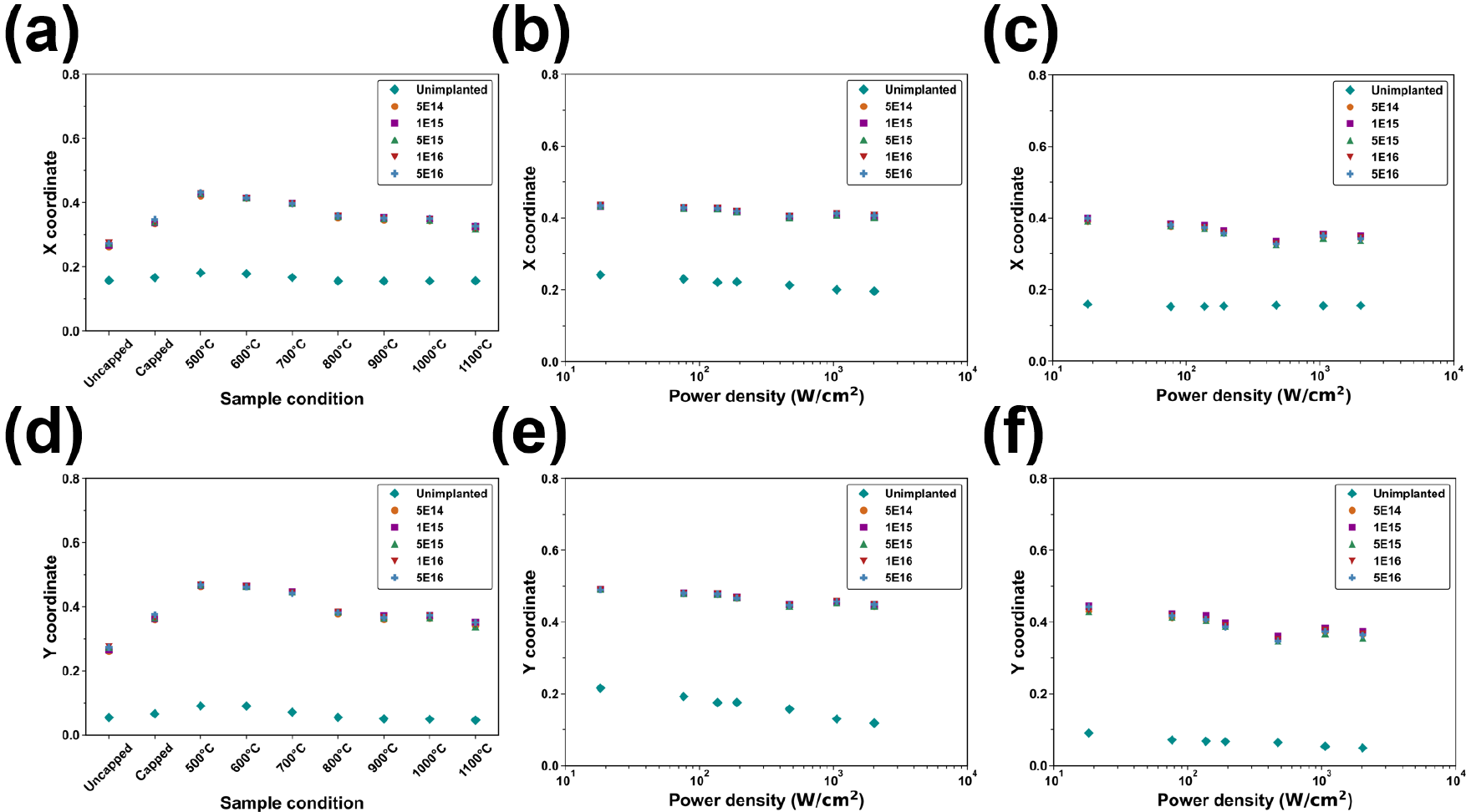}
      \caption{The evolution of x coordinate obtained using (a) 2021.27 $\text{W/cm}^2$ 325 nm excitation as a function of annealing temperature, and as a function of excitation power density following annealing at (b) 500 °C and (c) 1000 °C, and y coordinate obtained using (d) 2021.27 $\text{W/cm}^2$ 325 nm excitation as a function of annealing temperature, and as a function of excitation power density following annealing at (e) 500 °C and (f) 1000 °C.}
      \label{figureS_CIE_overall}
    \end{figure}

Figures \ref{figureS_CIE_unimplanted} to \ref{figureS_CIE_5E16} respectively illustrate the chromaticity coordinates extracted for the unimplanted, the 5E14 ions cm$^{-2}$, the 1E15 ions cm$^{-2}$, the 1E16 ions cm$^{-2}$, and the 5E16 ions cm$^{-2}$ In implanted areas across process stages and under varying power densities. Figure \ref{figureS_CIE_overall} illustrates the chromaticity coordinates ($x, y$) for all of the samples following each experimental stage, and as a function of excitation power density.

\section{Time-of-flight secondary ion mass spectrometry of implanted areas}
\label{SectionS5}

\begin{figure}[htbp]
      \centering
      \includegraphics[width=1\textwidth]{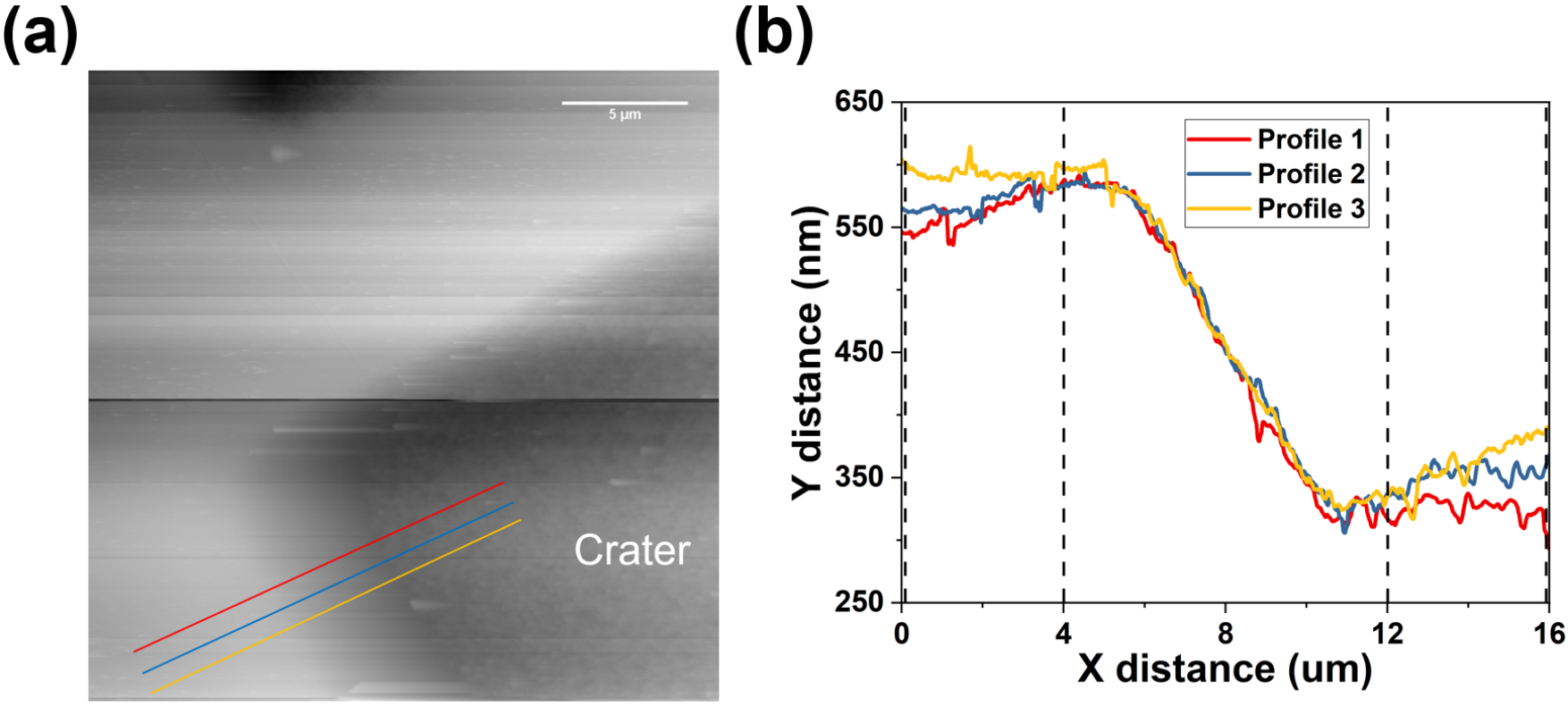}
      \caption{(a) The AFM topography of the 5E15 ions cm$^{-2}$ In implanted ToF-SIMS crater boundary with three cross-sectional lines. (b) Corresponding depth profiles colour-coded to match the lines in (a).}
      \label{figureS_5e15_AFM_image}
    \end{figure}

Figure \ref{figureS_5e15_AFM_image} shows the AFM topography and the height profile of the 5E15 ions cm$^{-2}$ In-implanted ToF-SIMS crater boundary. The crater depth was quantified by extracting three independent cross-sectional profiles across the crater edge. As indicated by the vertical dashed lines in Figure \ref{figureS_5e15_AFM_image}(b), two reference regions representing the exterior and interior of the crater were selected to compute the average y-value for each profile. The crater depth was determined to be $\sim$229.0 ±5.20 nm, derived from the average of the three independent profiles, with the error quoted as the standard error of the mean.

\begin{figure}[htbp]
      \centering
      \includegraphics[width=1\textwidth]{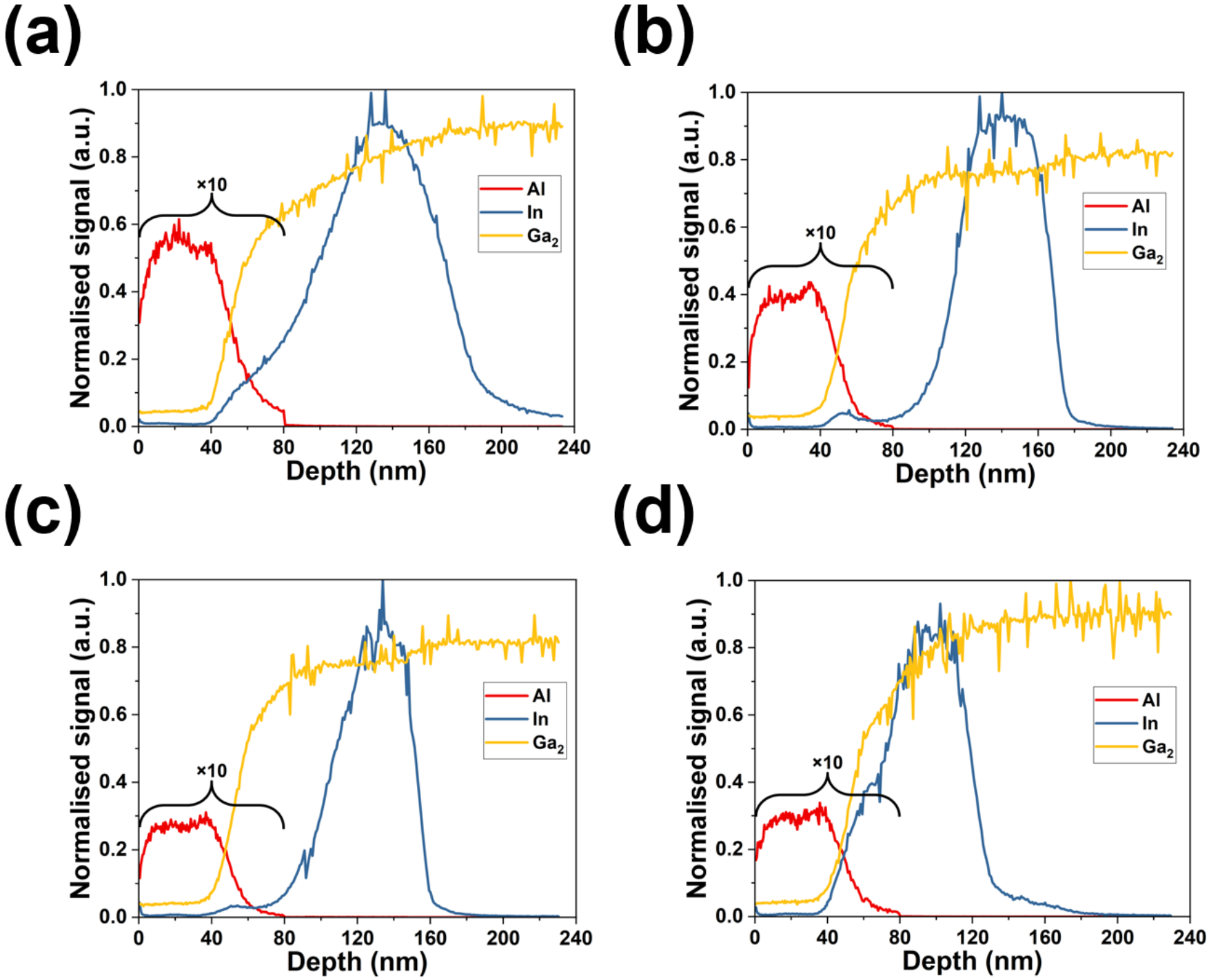}
      \caption{The ToF-SIMS profiles of (a) the 5E14 ions cm$^{-2}$, (b) the 1E15 ions cm$^{-2}$, (c) the 1E16 ions cm$^{-2}$, and (d) the 5E16 ions cm$^{-2}$ In implanted area. All profiles were obtained using an ion current of 10 pA and initial signals of the Al element have been enlarged by a factor of 10 for clarity.}
      \label{figureS_ToF_SIMS}
    \end{figure}

Figures \ref{figureS_ToF_SIMS}(a) to \ref{figureS_ToF_SIMS}(d) respectively presents the elemental depth profile of the 5E14 ions cm$^{-2}$, the 1E15 ions cm$^{-2}$, the 1E16 ions cm$^{-2}$, and the 5E16 ions cm$^{-2}$ In implanted areas.

\end{document}